\documentclass[11pt]{article}
\usepackage[margin=1in]{geometry}
\usepackage{amsmath,amssymb,amsthm}
\usepackage{booktabs,graphicx,caption,subcaption,threeparttable,array,multirow}
\usepackage[hidelinks]{hyperref}
\usepackage[round]{natbib}
\usepackage[T1]{fontenc}\usepackage{mathptmx}\usepackage[protrusion=false,expansion=false]{microtype}
\usepackage{setspace}
\usepackage{placeins}

\newtheorem{proposition}{Proposition}
\newcommand{\inhouse}{\text{in-house}}
\newcommand{\tabnote}[1]{\par\vspace{3pt}\parbox{\textwidth}{\footnotesize #1}}
\title{\Large AI, Specialization, and the Organization of Knowledge Work:\\
\large Evidence from Legal Services}
\author{
    Liu Zhang\thanks{\texttt{liu@micro1.ai}}\\
    \small micro1
    \and
    Mark Esposito\thanks{\texttt{mark@micro1.ai}}\\
    \small micro1\\
    \small Northeastern University
    \and
    Isabel Yang\thanks{\texttt{isabel@micro1.ai}. Led the development of the micro1 legal benchmarks cited in this paper.}\\
    \small micro1
}

\date{November 2026}

\begin{document}
\maketitle
\begingroup\renewcommand{\thefootnote}{}\footnotetext{\textit{Disclosure.} The authors are employed by micro1, which produces and publishes three of the legal-AI benchmarks cited in Sections~\ref{sec:data} and~\ref{sec:capability} (LegalReasonBench, Ivo-micro1 ContractBench, Crosby-micro1 RedlineBench); Isabel Yang led their development. All benchmark figures are the publicly released results; the empirical findings on courts, employment, spending, and firm revenue do not depend on them, and Section~\ref{sec:capability} reaches the same conclusion on the independent public benchmarks alone. The authors have no other financial interests related to this work.}\addtocounter{footnote}{-1}\endgroup

\begin{abstract}
\noindent Artificial intelligence makes specialized knowledge cheaper to use, but its effects on professional-service firms depend on how clients reallocate work and how suppliers respond. We develop a framework linking individual self-provision, enterprise make-or-buy decisions, and pricing by specialist firms. As AI improves lower-priced alternatives, clients may reallocate work across outside providers without bringing it in-house. If the departing work is more price-sensitive than the work that remains, premium firms may raise their optimal markups even as their share of work declines. We examine these margins in U.S. legal services. Across 89 federal districts, a difference-in-differences analysis finds a 6.6-percentage-point increase in non-prisoner self-representation relative to prisoner petitions by fiscal 2025, remaining substantial after adjusting for pre-existing trends. In contrast, lawyer employment shows no clear post-2022 shift toward client firms. Legal-AI benchmarks reveal continuing limitations in reliably completing complex, multi-step work, offering one possible explanation. Meanwhile, the largest law firms reported 13\% revenue growth in 2025 despite nearly flat billable hours, while lower-priced firms experienced stronger demand growth. Reported AI integration is also stronger among firms with lower revenue per lawyer. These patterns highlight the importance of changes in work allocation and supplier competition, even as much of the underlying reallocation predates generative AI. The findings illustrate why cheaper expertise need not produce either widespread internalization or declining prices for premium professional services.
\end{abstract}

\vspace{0.5em}
\noindent\textbf{Keywords:} generative AI,
firm boundaries, knowledge specialization, make-or-buy decisions, legal services, professional services, pro se litigation, mobile demand

\vspace{0.3em}
\noindent\textbf{JEL Classification:}
D23, L22, O33, L84, K41

\newpage
\section{Introduction}

AI is dramatically lowering the cost of acquiring specialized knowledge. Increasingly, it can also perform professional tasks directly from start to finish. These advances raise a fundamental question in the economics of firm boundaries: as expertise becomes cheaper, which activities will clients perform themselves, and which will they continue to purchase from specialist firms? This choice depends on how AI changes the relative costs and capabilities of clients and outside providers. Yet the make-or-buy decision is only one margin of adjustment. AI may also improve lower-priced providers' ability to compete for work, allowing clients to reallocate purchases across outside suppliers without bringing production in-house. The resulting changes may therefore involve both firm boundaries and the composition and pricing of specialist services.

These questions connect two strands of organizational economics. The theory of the firm examines why activities are organized within firms rather than through markets \citep{coase1937,grossmanhart1986}, while the economics of knowledge studies how the costs of acquiring and communicating expertise shape specialization and the division of labor \citep{garicano2000,garicanorossi2006}. Technological change can shift organizational boundaries in different directions \citep{bakerhubbard2003}, and specialization and transaction costs help determine the scope of law firms \citep{garicanohubbard2009}. More recently, \citet{garicano2026} argues that AI may erode specialists' cost advantage and shift production toward client organizations. We build on this literature by examining not only the boundary between clients and specialist providers, but also how cheaper expertise changes competition and the allocation of work among competing specialist providers.

Legal services provide a useful setting for studying these adjustments. Corporate legal departments routinely allocate work among in-house lawyers, lower-priced outside providers, and premium law firms, depending on the nature of the work and the capabilities of each supplier. This disaggregation of legal work, and the movement of demand across law-firm segments, predates generative AI \citep{reganheenan2010,acc2025,engelland2026}. AI may accelerate these changes by expanding the range of tasks that clients and lower-priced providers can perform, while leaving some advantages of premium firms, such as specialized judgment and client relationships, intact. A client may consequently retain its relationship with a premium firm while assigning a different mix of work to that firm. The resulting change in the allocation of work can also change the value of the expertise that remains in demand \citep{autorthompson2025}.

This reallocation also has implications for specialist pricing. Premium firms serve clients and matters that differ in how readily they can be assigned to alternative providers. As AI improves these alternatives, more substitutable work may move elsewhere, while clients continue to rely on premium firms for matters requiring specialized judgment, reputation, or relationship-specific knowledge. If the departing work is more price-sensitive than the work that remains, residual demand becomes less elastic, and the premium firm's optimal markup may rise even as its volume declines \citep{franksalkever1997,chenriordan2008}. Changes in the composition of work can also affect average realized rates, allowing premium firms to report rising rates even as some work shifts to lower-priced providers.

We formalize these ideas through three linked decisions. Individuals choose whether to hire counsel, represent themselves, or forgo a legal claim. Enterprises decide which recurring activities to perform internally, weighing the fixed cost of maintaining the necessary capabilities against the cost of outsourcing. Conditional on
outsourcing, clients choose among differentiated providers and
determine whether AI use is permitted. Outside suppliers choose
their AI adoption and pricing strategies in response to these
constraints. The model shows how client restrictions can limit
the returns to AI adoption, while improvements in lower-priced
alternatives can shift contestable work away from premium firms.
Selection into the remaining client base may then increase
premium markups even as the firm's volume declines. The framework identifies the conditions under which self-provision, internalization, and substitution among outside suppliers become attractive. It also shows why cheaper expertise need not lead to lower premium prices or a uniform contraction of specialist firms.

We examine these mechanisms using data from the U.S. legal-services market. Federal court records reveal changes in self-representation across districts and case types. Employment and spending data track the organization and demand for legal services, while law-firm revenue, billable hours, and practice-level series capture differences across outside providers. We also examine measures of AI integration and model capabilities. Legal work varies considerably in recurrence, standardization, judgment requirements, and the availability of alternative suppliers, providing a setting in which the different margins of adjustment can be examined together.

The empirical evidence reveals different patterns across these margins. First, self-representation in federal civil litigation has risen sharply since ChatGPT launched in 2022. The share of non-prisoner cases involving a self-represented party reached 16.8\% in fiscal 2025, after remaining between 9.5\% and 12.1\% for nearly two decades. Using a difference-in-differences design comparing non-prisoner cases with prisoner petitions across 89 federal districts, we estimate a 6.6-percentage-point relative increase by 2025. The finding remains substantial at 5.7 points after accounting for a modest upward pre-existing trend. Increases are particularly pronounced in several case types requiring substantial drafting, consistent with AI reducing the cost of self-representation. Meanwhile, real household spending on legal services remained above its pre-pandemic trend, indicating that greater self-representation has not been accompanied by a broad decline in paid legal demand.

Second, we find little evidence of a new shift in lawyer employment toward client firms. The share of privately employed lawyers working outside the legal-services industry rose substantially during the 2010s, before generative AI, but stood at 29.0\% in 2024, close to 28.7\% in 2019. 
Within-state estimates show no clear post-2022 acceleration in in-house hiring, nor is industry-level AI adoption positively associated with such hiring. Evidence on AI capabilities offers one possible explanation. Model performance on legal reasoning and research tasks has improved substantially \citep{guha2023,vals2026}, but reliability remains limited on more demanding, multi-step legal work \citep{harvey2026,micro1lrb,micro1contract}. In an October 2026 evaluation of Harvey's Legal Agent Benchmark conducted by Vals AI, the leading model fully passed only about one-quarter of tasks \citep{vals2026}. Separate micro1 evaluations similarly document substantial limitations in long-horizon legal reasoning and context-sensitive contract review \citep{micro1lrb,micro1contract}. AI may therefore improve the execution of individual tasks before it changes the relative costs of fulfilling more complex legal demand in-house. 

Third, the outside-provider market exhibits changes in demand and pricing consistent with our model predictions. In 2025, the largest U.S. law firms reported 13\% revenue growth despite nearly flat billable hours, while lower-priced firms experienced stronger demand growth. Realized rates rose substantially, and midsize firms experienced stronger hours growth. 
Comparing practice-level demand with legal-agent benchmark results, we find no systematic evidence that specialties with stronger AI capabilities experienced faster declines after 2022. Instead, several of these specialties, particularly patent and tax, had already experienced declining billable hours before generative AI became widely available. 
Meanwhile, firms with lower revenue per lawyer report stronger AI integration, consistent with AI improving clients' alternatives to premium providers. Together, these patterns suggest that changes in the allocation of work among outside suppliers may be an important margin of adjustment.

The paper contributes to the economics of knowledge organization in two ways. Conceptually, it shows why cheaper knowledge need not imply a uniform contraction of specialist suppliers: clients reallocate different kinds of work, and selection into the remaining customer base can change specialist markups. 
Empirically, we examine three margins of adjustment in the legal-services market within a common framework. We extend recent evidence on rising self-representation in federal civil litigation \citep{shahlevy2026,cohensasson2026} by comparing changes in non-prisoner filings with prisoner petitions within the same districts. We then connect these patterns to evidence on in-house legal employment, demand across law-firm segments, and AI integration among outside providers. Together, these findings provide a broader picture of how the organization of legal work has evolved alongside the diffusion of AI, highlighting distinct adjustments in self-provision, internalization, and competition among specialist suppliers.

Section~\ref{sec:model} develops the framework. Sections~\ref{sec:data} and \ref{sec:spec} describe the data and empirical designs. Section~\ref{sec:results} examines the household, enterprise, and outside-provider margins; Section~\ref{sec:robust} reports sensitivity analyses. Section~\ref{sec:capability} relates the evidence to the state of legal AI capability. Section~\ref{sec:conclusion} concludes.

\section{A model of the organization and pricing of legal work}\label{sec:model}

Cheaper expertise does not necessarily imply that clients bring more work
in-house or that suppliers charge lower prices. The effect depends
on the decisions being made. Individuals may perform legal work themselves
rather than hire counsel. Enterprises may internalize recurring work, but
only when the cost of maintaining the relevant capability is justified by
its use. And when work remains outsourced, clients may redirect some tasks
from premium firms to lower-priced providers. Such reallocation depends
not only on supplier capabilities but also on clients' willingness to
accept AI-assisted work, particularly where quality is difficult to verify.

We formalize these three margins in turn: household self-provision,
enterprise make-or-buy, and the allocation and pricing of outsourced work.
The first two problems take the cost of outside counsel as given. The
third examines supplier adoption and pricing conditional on a
pool of outsourced matters. The framework isolates the different margins
without claiming to solve all markets simultaneously.

\subsection{Households: hire, self-provide, or forgo}

An individual has a legal problem worth $v\geq0$ if resolved
successfully. Hiring counsel costs $p_c$ and succeeds with probability
$\pi_L$. Self-representation succeeds with probability $\pi_S$,
where $0<\pi_S<\pi_L$, and requires drafting effort with monetary-
equivalent cost $d>0$. Not pursuing the matter yields zero. Consumer
AI reduces the drafting cost to $(1-a^H)d$, where $a^H\in[0,1]$
measures the proportional reduction. Holding the success probabilities
fixed isolates the drafting-cost channel.

The individual chooses the option with the greatest expected payoff:
\begin{equation}
 V(v,a^H)=\max\!\left\{
 \underbrace{\pi_L v-p_c}_{\text{hire counsel}},\quad
 \underbrace{\pi_S v-(1-a^H)d}_{\text{self-represent}},\quad
 \underbrace{0}_{\text{do not proceed}}
 \right\}.
 \label{eq:household_objective}
\end{equation}

Self-representation is optimal if its payoff exceeds both alternatives:
\begin{align}
 \pi_S v &\geq (1-a^H)d,
 \label{eq:self_participate}\\
 p_c-(1-a^H)d &\geq (\pi_L-\pi_S)v.
 \label{eq:self_substitute}
\end{align}

Equivalently, when the interval is nonempty, self-representation
is chosen for
\begin{equation}
 \underbrace{\frac{(1-a^H)d}{\pi_S}}_{\underline v(a^H)}
 \leq v\leq
 \underbrace{\frac{p_c-(1-a^H)d}{\pi_L-\pi_S}}
             _{\overline v(a^H)}.
 \label{eq:self_interval}
\end{equation}

An increase in $a^H$ lowers $\underline v$ and raises $\overline v$.
It can thus induce participation among individuals who would otherwise
forgo the matter and substitution among individuals who would
otherwise hire counsel. Both effects can raise self-represented
filings. Because some new filings come from people who otherwise
would not proceed, increased self-representation need not imply
a one-for-one decline in represented cases.

The mechanism is most relevant when drafting is a material part of
the individual's cost; it is weaker for short-form proceedings or
where AI is inaccessible.

\subsection{Enterprises: the make-or-buy boundary}

An enterprise requires $q>0$ units per year of a recurring category of
legal work. Outsourcing costs $p_o$ per unit. Internal production costs
$c$ per unit plus a fixed annual cost $F>0$ of maintaining the required
knowledge, oversight, and coordination. The enterprise chooses
$z\in\{0,1\}$, where $z=1$ denotes internal production:
\begin{equation}
 \min_{z\in\{0,1\}}\{z(F+qc)+(1-z)qp_o\}
 =\min\{F+qc,\,qp_o\}.
 \label{eq:enterprise_objective}
\end{equation}
If $p_o>c$, the firm internalizes the category if and only if
\begin{equation}
 q\geq q^*\equiv\frac{F}{p_o-c}.
 \label{eq:threshold}
\end{equation}
Otherwise, with $F>0$, outsourcing is preferred. The threshold applies
to a category of work, not to an entire legal matter or a full legal
department. The client may internalize one category and
continue to purchase other work. AI can lower $F$ and $c$, but AI-enabled outside suppliers
can also lower $p_o$. Letting these quantities depend on technological
progress $t$ yields
\begin{equation}
 \frac{dq^*}{dt}=
 \frac{F_t(p_o-c)-F(p_{o,t}-c_t)}{(p_o-c)^2}.
 \label{eq:threshold_cs}
\end{equation}
A fall in $F$ or $c$, holding the outsourcing price fixed, makes
internalization more attractive; a fall in $p_o$, holding internal
costs fixed, makes outsourcing more attractive. The net change
in the firm boundary is therefore ambiguous. 
Moreover, lower internal task costs need not immediately change legal-department
employment if existing in-house lawyers can
perform more work without increasing headcount. Here $p_o$ summarizes
the cost of an outside solution, possibly involving several providers.
Conditional on the enterprise buying rather than making, the next
subsection examines which outside supplier receives the work and how
supplier adoption affects its price.

\subsection{Premium suppliers: client restrictions and endogenous AI adoption}

\paragraph{Why premium suppliers exist.}
Premium law firms have advantages in specialized judgment, information
accumulated across transactions, reputation, and client relationships.
These assets can be valuable when quality is hard to verify and mistakes
are costly \citep{defontenay2015,uzzilancaster2004,levintadelis2005}.
We represent this differentiation by the probabilities of a
satisfactory outcome, $k_H>k_L$, on the relevant set of matters,
together with a client-specific match advantage $\eta_i$ for the
premium firm, which can capture prior relationships and matter-specific
fit. The premium firm need not be better at every task. Client $i$ has
a legal matter worth $v_i$ if resolved satisfactorily. Its expected
incremental value from premium provision is $v_i(k_H-k_L)$. Absent AI,
the client prefers $H$ to $L$ when
\begin{equation}
 v_i(k_H-k_L)+\eta_i\geq p-r,
 \label{eq:quality_sort}
\end{equation}
where $p$ is the premium firm's price and $r$ is the challenger's price.
Holding the quality gap fixed, the greater the value at stake, the more
the client is willing to pay for the premium supplier's quality advantage.

\paragraph{Client permission.}
Adopting AI need not mean using it on every legal matter for all clients.
Clients may restrict AI use through outside-counsel
guidelines or matter-specific instructions, while law firms may
require client consent before using AI with confidential
information \citep{trgenai2025,aon2024}. These restrictions
limit the work on which firms can realize the benefits of AI
integration and may therefore help explain differences in
adoption incentives across suppliers.

We model permission at the legal-matter level: the same client
may authorize AI on some matters but prohibit it on others.
Let $i$ index legal matters and $v_i$ denote the value at
stake in matter $i$. Suppose permission decisions are made
before firms choose their AI integration levels. Allowing AI
provides an anticipated benefit $b>0$, including potential
improvements in efficiency and service quality, but entails
a perceived expected loss $\rho v_i$, where $\rho>0$ captures
concerns about reliability, confidentiality, or other risks
whose consequences increase with the value of the matter.
AI is permitted when
\begin{equation}
 A_i=\mathbf{1}\{b-\rho v_i\geq0\},
 \qquad
 v_i\leq\bar v\equiv\frac{b}{\rho}.
 \label{eq:ai_permission}
\end{equation}

We classify matters with $A_i=0$ as \emph{restricted} ($B$)
and those with $A_i=1$ as \emph{flexible} ($F$).
Because the value of a matter also affects willingness to pay
for premium legal services, higher-value matters may be both
more likely to use premium providers and less likely to permit
AI.

\paragraph{Adoption and allocation of outsourced work.}
Each supplier chooses $x_j\in\{0,1\}$ before competing for work,
paying a fixed cost $K_j>0$ if it adopts. An adopting supplier can
use AI only on flexible matters. For these matters, adoption saves
$s_j\geq0$ in marginal production cost and improves the value of
the supplier's services by $\delta_j\geq0$. We impose neither
$s_L>s_H$ nor $\delta_L>\delta_H$: the relative benefits of
adoption are not assumed to favor challengers.

Conditional on outsourcing, supplier adoption, and AI permission,
the client assigns matter $i$ to the provider offering the
greater net value:
\begin{equation}
 j_i^*\in\arg\max_{j\in\{H,L\}}
 \left\{v_i k_j+\eta_i\mathbf{1}\{j=H\}
       -p_j+\delta_j x_j A_i\right\},
 \qquad p_H=p,\quad p_L=r.
 \label{eq:task_objective}
\end{equation}
The matter is assigned to the premium provider when
\begin{equation}
 p+\xi A_i\leq r+v_i(k_H-k_L)+\eta_i,
 \qquad \xi\equiv\delta_Lx_L-\delta_Hx_H.
 \label{eq:assign}
\end{equation}
Thus, adoption affects the allocation of matters directly only
where AI use is permitted. Restricted matters may still move
between providers as fees change, even though neither provider
can use AI on them.

\paragraph{Demand for premium services.}
Consider a pool of outsourced legal matters, classified as
restricted ($B$) or flexible ($F$) according to whether AI
use is permitted. For tractability, we hold the challenger's
fee $r$ fixed and consider equilibria in which the premium
firm charges a higher price ($p^*>r$).

Matter $i$ has reservation value for premium provision
\[
w_i=r+v_i(k_H-k_L)+\eta_i.
\]
Conditional on permission status, assume that reservation
values have exponential survival functions over the relevant
price range. Restricted matters are less price-sensitive,
$0<\lambda_B<\lambda_F$. This ordering is an additional
assumption, not a direct implication of the AI permission
threshold.

Let $b_B,b_F>0$ denote the respective masses of restricted
and flexible matters per unit of market activity. Premium
demand is
\begin{align}
 D_B(p)&=b_Be^{-\lambda_Bp},\nonumber\\
 D_F(p,\xi)&=b_Fe^{-\lambda_F(p+\xi)},\nonumber\\
 D_H(p,\xi)&=D_B(p)+D_F(p,\xi).
 \label{eq:demand}
\end{align}
The baseline effect of $r$ is absorbed into the
reservation-value distributions. AI adoption affects demand
directly only for flexible matters. We restrict attention
to parameters with $p+\xi\geq0$ at the relevant optima,
so demand does not exceed the available mass of matters.

With $M>0$ outsourced matters, the quantities supplied by
the premium firm ($H$) and challenger ($L$) are
\begin{equation}
 Q_H^g=MD_g,\qquad Q_L^g=M(b_g-D_g),
 \qquad g\in\{B,F\}.
 \label{eq:outside_quantities}
\end{equation}
Restricted matters can still move between providers as
prices change, but neither provider can use AI on them.

\paragraph{Premium pricing.}
Let $m_H$ denote the premium firm's marginal production cost without
AI. After adoption, its flexible work costs $m_H-s_Hx_H$ per task;
restricted work still costs $m_H$. For a given adoption profile, the
premium firm chooses a common price $p$ to maximize operating profit:
\begin{equation}
 p^*(x_H,x_L)\in\arg\max_{p\geq m_H}
 M\big[(p-m_H)D_B(p)
 +(p-m_H+s_Hx_H)D_F(p,\xi)\big].
 \label{eq:premium_objective}
\end{equation}
At an interior optimum, its first-order condition is
\begin{equation}
 \underbrace{D_B\big[1-\lambda_B(p-m_H)\big]
 +D_F\big[1-\lambda_F(p-m_H+s_Hx_H)\big]}_{G(p,\xi,x_H)}=0.
 \label{eq:premium_foc}
\end{equation}
Equivalently, the markup over the non-AI marginal cost satisfies
\begin{equation}
 p^*-m_H=
 \frac{D_B+D_F-\lambda_Fs_Hx_HD_F}
 {\lambda_BD_B+\lambda_FD_F}.
 \label{eq:markup}
\end{equation}
Without premium-firm adoption this reduces to the familiar inverse
weighted elasticity, $p^*-m_H=1/\bar\lambda$, with
$\bar\lambda=(\lambda_BD_B+\lambda_FD_F)/(D_B+D_F)$.
The second-order condition at the relevant optimum is
\begin{equation}
 G_p=-\lambda_BD_B[2-\lambda_B(p-m_H)]
 -\lambda_FD_F[2-\lambda_F(p-m_H+s_Hx_H)]<0.
 \label{eq:premium_soc}
\end{equation}

\begin{proposition}\label{prop:price}
Hold the premium firm's adoption decision fixed. Suppose both client
groups have positive mass, $\lambda_F>\lambda_B$, and the pricing
optimum is interior and stable ($G_p<0$). An improvement in the
challenger's effective offering to flexible clients (an increase in
$\xi$) raises the premium firm's optimal price and reduces its total
volume per unit of market activity:
\[
 \frac{\partial p^*}{\partial\xi}>0,
 \qquad
 \frac{dD_H(p^*(\xi),\xi)}{d\xi}<0.
\]
\end{proposition}

The result follows directly from the first-order condition. Writing
$u=p^*-m_H\geq0$ and $h=s_Hx_H\geq0$, equation
\eqref{eq:premium_foc} implies
$u\leq1/\bar\lambda<1/\lambda_B$. Hence the restricted term in
\eqref{eq:premium_foc} is positive and the flexible term is negative.
Therefore
\begin{equation}
 G_{\xi}=-\lambda_FD_F[1-\lambda_F(u+h)]>0,
 \qquad
 \frac{\partial p^*}{\partial\xi}=-\frac{G_{\xi}}{G_p}>0,
 \label{eq:price_selection_derivatives}
\end{equation}
and
\begin{equation}
 \frac{dD_H}{d\xi}=-\lambda_FD_F+D_{H,p}
 \frac{\partial p^*}{\partial\xi}<0.
 \label{eq:price_selection_cs}
\end{equation}
Better challenger services remove relatively more flexible,
price-sensitive work, leaving a less elastic premium customer mix.
This is a conditional \emph{selection} result, not a claim that AI
always raises premium prices. Premium-firm adoption lowers costs on
flexible work and improves the firm's own offering, creating opposing
price effects when both firms adopt. Nor does the result require the
premium firm to lose all of its flexible clients.

\paragraph{Adoption incentives.}
Let $V_H(x_H,x_L)$ be the maximized operating profit in
\eqref{eq:premium_objective}, before adoption cost. At the premium
firm's optimal price, the challenger's operating profit is
\begin{equation}
 V_L(x_H,x_L)=
 (r-m_L)\big[Q_L^B+Q_L^F\big]+s_Lx_LQ_L^F,
 \label{eq:challenger_profit}
\end{equation}
where $m_L$ is its marginal cost without AI. The cost saving applies
only to its authorized flexible work. Each firm chooses adoption to
maximize $V_j(x_H,x_L)-K_jx_j$. In a pure-strategy adoption equilibrium,
firm $j$ adopts when its incremental operating profit covers $K_j$:
\begin{equation}
 \begin{aligned}
 V_H(1,x_L)-V_H(0,x_L)&\geq K_H,\\
 V_L(x_H,1)-V_L(x_H,0)&\geq K_L.
 \end{aligned}
 \label{eq:adoption_condition}
\end{equation}

The challenger can gain both new business and lower production costs.
For example, holding $x_H$ fixed, its incremental operating profit
from adopting is
\begin{equation}
 V_L(x_H,1)-V_L(x_H,0)=
 (r-m_L)\big[\Delta Q_L^B+\Delta Q_L^F\big]
 +s_L Q_L^{F,1},
 \label{eq:challenger_adoption}
\end{equation}
where $\Delta Q_L^g=Q_L^g(x_H,1)-Q_L^g(x_H,0)$.
The restricted segment may also respond indirectly if the premium
firm changes its price, even though AI is not used on those matters.
The premium firm's adoption payoff similarly reflects cost savings
and the ability to retain or attract flexible work, not savings on
restricted matters. A firm with a large restricted client base may
therefore have a relatively low return to deploying AI in client work,
even if it is large and well resourced. This is not an unconditional
prediction that lower-priced firms adopt more: adoption also depends
on $K_j$, $s_j$, $\delta_j$, and the size of each firm's contestable
business. Firm-level adoption and authorized use on specific matters
are different outcomes.

\paragraph{Realized rates and aggregate demand.}
The common-price assumption delivers the within-service markup result.
If firms instead price the two client groups separately, changes in
work composition can raise their average realized rates even without
higher prices for either group. With $p_B>p_F$ fixed,
\begin{equation}
 \bar p_H=\frac{p_BD_B+p_FD_F}{D_B+D_F},
 \qquad
 \frac{\partial\bar p_H}{\partial D_F}=
 \frac{(p_F-p_B)D_B}{(D_B+D_F)^2}<0.
 \label{eq:composition_price}
\end{equation}
Thus a decline in flexible work raises the average realized rate
through composition alone. More generally, if $h$ denotes outsourced
hours per unit of activity, $s$ the premium firm's share of outside
hours, and $\bar r$ its average realized rate, then
\begin{equation}
 H=Mhs,\qquad R=\bar r H,\qquad
 \Delta\ln R=\Delta\ln M+\Delta\ln h+
 \Delta\ln s+\Delta\ln\bar r.
 \label{eq:decomp}
\end{equation}
Market growth can sustain premium-firm hours even as its share falls.
Aggregate realized rates cannot separate within-service markup changes
from shifts in the types of matters or lawyers billed.

\subsection{Implications for the data}

The model identifies three margins along which AI can change the allocation of legal work. Lower drafting costs
can increase individual self-representation and participation. For
enterprise clients, internalization depends on the relative costs of
maintaining a recurring capability and buying the same work outside;
AI can benefit both sides of that comparison. Among outside providers,
clients with the greatest willingness to pay for trusted expertise
may also be the least willing to authorize AI. This can limit the
premium firm's immediate return to AI deployment while giving other
providers an opportunity to win flexible work. Under the stated
elasticity and pricing assumptions, that reallocation can increase
the premium firm's optimal price; work composition alone can also
raise its realized rate.

The adoption and permission decisions are explicit, but the analysis
remains partial equilibrium: the challenger's fee, the total pool of
outsourced work, and the distribution of client characteristics are
taken as given. We do not claim that the high-willingness-to-pay clients
are empirically more likely to prohibit AI; that correlation is a
mechanism to test. Nor do the data on law-firm integration identify
which clients authorized AI or distinguish internal tool access from
client-facing deployment.

\section{Data}\label{sec:data}

Table~\ref{tab:summary} summarizes the main data sources. Court records, ACS microdata, and government spending series are public; law-firm aggregates, survey rankings, and published benchmarks come from the cited providers. Construction details and source definitions appear in Appendix~\ref{app:data}.

\paragraph{Federal courts.} The Administrative Office's Table C-13 reports, for each district and fiscal year, civil cases with and without a self-represented party, separately for prisoner petitions and all other cases. We transcribed FY2005--2025. The Northern District of Florida carried the 3M earplug multidistrict litigation, several hundred thousand largely self-represented short-form complaints that were 90\% of national non-prisoner pro se filings in FY2020; it and the four territorial districts are excluded, leaving 89 districts whose series we call the main series. The district panel runs FY2010--2025: 89 districts $\times$ two groups $\times$ 16 years, with the share of cases with a self-represented party as the outcome. Prisoner petitions in FY2016 include a one-year wave of motions to vacate filed with appointed counsel after \emph{Welch v.\ United States} made \emph{Johnson} retroactive; that year is dropped from the estimates and shown in the figures. The Federal Judicial Center's Integrated Database supplies 4.9 million civil records filed FY2010--2025 with nature-of-suit codes and a pro se indicator; we use it for the case-type decomposition and for an exclusion based on recorded multidistrict-litigation status rather than on district.

\paragraph{Lawyer employment.} The ACS one-year public-use microdata identify lawyers (occupation 2100, which is defined identically across the 2010, 2014, and 2018 code vintages) and their industry. In-house lawyers are private-sector employees outside legal services (industry 7270); law-firm lawyers are private-sector employees in legal services. Restricting both groups to employees removes partners and the self-employed from both sides and makes the comparison like for like; Appendix~\ref{app:data} reports the version that includes them. The balanced panel covers 42 states with positive cells in every year, 2010--2024 (no 2020 one-year file). Person weights give employment; replicate weights give survey errors.

\paragraph{Spending and revenue.} BEA personal consumption of legal services and its price index give real household spending, 2000--2025. The Census Quarterly Services Survey gives legal-services revenue by class of customer, deflated by the BLS producer price index for legal services; the business series runs from 2000 with gaps before 2012 and is continuous 2012--2025, and the household series has gaps. The American Lawyer gives Am Law 100 gross revenue 2010--2025, deflated by the same index. Thomson Reuters' annual reports give large-firm billable-hours growth 2007--2025, practice-level growth for nine practices 2019--2025, segment detail for 2025, and quarterly practice growth from its Law Firm Financial Index, continuous from 2022Q3.

\paragraph{Practices, capability, and firms.} Model capability by practice area comes from an independent full run of Harvey's Legal Agent Benchmark on 1,251 matters across 24 areas \citep{gengliu2026}, taken as criterion accuracy for Claude Sonnet 4.6 in Claude Code, the strongest plain-model baseline in that run, and mapped to Thomson Reuters' practices. Capability over time comes from Vals AI's public leaderboards for LegalBench \citep{guha2023}, the Legal Research Bench, and the Legal Agent Benchmark, and from three expert-authored benchmarks published by micro1 in 2026 that grade complete enterprise work product with partial credit:\footnote{These benchmarks are produced by micro1, the authors' employer, and one author led their development; see the disclosure on the title page. They are used as published and alongside independent benchmarks, not in place of them.} Crosby-micro1 RedlineBench (June), multi-turn SaaS negotiations scored on attorney rubrics \citep{micro1redline}; LegalReasonBench (preliminary results in May, updated results in October), 50 long-horizon matters in 122 assignments scored on IRAC-weighted rubrics \citep{micro1lrb}; and Ivo-micro1 ContractBench (October), 350 contract-review tasks across five agreement types scored against five attorneys' redlines \citep{micro1contract}. ContractBench's published scores by agreement type give a second, task-specific reading of capability for three of the nine practices. Firm-level data are Vault's Technology \& Innovation score (associates' 1--10 rating of their firm's technology, top 30 firms per edition, editions 2020--2027), Chambers Associate's 2026 AI-integration tiers for 64 firms, and Am Law FY2024 revenue per lawyer.

\begin{table}[tbp]\centering\caption{Summary statistics of the series used in the main text}\label{tab:summary}
\footnotesize
\resizebox{\textwidth}{!}{\begin{tabular}{llllrrrrr}
\toprule
 & Variable & Unit & Period & $N$ & Mean & SD & Min & Max \\
\midrule
\multicolumn{9}{l}{\textit{A. Federal civil courts}} \\
 & Pro se share, non-prisoner cases & district-FY, \% & FY2010--25 & 1,424 & 14.4 & 7.3 & 0.2 & 82.4 \\
 & Pro se share, prisoner petitions & district-FY, \% & FY2010--25 & 1,424 & 90.1 & 10.7 & 18.2 & 100.0 \\
 & Cases per district-FY, non-prisoner & district-FY, cases & FY2010--25 & 1,424 & 2,502 & 3,609 & 154 & 47,274 \\
 & Civil records per fiscal year & FY, records & FY2010--25 & 16 & 239,755 & 9,416 & 219,824 & 261,656 \\
\multicolumn{9}{l}{\textit{B. Lawyer employment}} \\
 & In-house lawyers (private employees) & state-year, persons & 2010--24 & 588 & 3,589 & 4,651 & 17 & 27,763 \\
 & Law-firm lawyers (private employees) & state-year, persons & 2010--24 & 588 & 10,032 & 12,624 & 136 & 73,050 \\
 & Unweighted records per in-house cell & state-year, records & 2010--24 & 588 & 37.9 & 48.8 & 1.0 & 303.0 \\
\multicolumn{9}{l}{\textit{C. Spending and revenue}} \\
 & Household consumption of legal services & year, \$bn 2017 & 2010--25 & 16 & 104.6 & 3.6 & 100.6 & 112.0 \\
 & Law-firm revenue, business customers & year, \$bn 2017 & 2000--25 (gaps) & 22 & 213.6 & 24.5 & 189.9 & 266.1 \\
 & Am Law 100 gross revenue & year, \$bn 2017 & 2010--25 & 16 & 97.0 & 12.0 & 82.8 & 120.1 \\
 & Large-firm billable hours growth & year, \% & 2007--25 & 19 & 0.4 & 2.1 & -5.1 & 4.1 \\
\multicolumn{9}{l}{\textit{D. Practices and capability}} \\
 & Practice demand growth, annual & practice-year, \% & 2019--25 & 61 & -0.2 & 3.9 & -10.2 & 10.6 \\
 & Practice demand growth, quarterly (4 practices) & practice-quarter, \% & 2022Q3--26Q3 & 65 & 1.0 & 2.4 & -10.3 & 4.9 \\
 & Model capability, criterion accuracy & LAB practice area, \% & 2026 & 24 & 82.9 & 4.3 & 74.5 & 90.8 \\
\multicolumn{9}{l}{\textit{E. Firms}} \\
 & Technology satisfaction score & firm-edition, 1--10 & 2020--27 eds. & 235 & 8.8 & 0.3 & 8.2 & 9.7 \\
 & Revenue per lawyer, Chambers-matched firms & firm, \$m & FY2024 & 41 & 1.4 & 0.4 & 0.7 & 2.3 \\
\bottomrule
\end{tabular}}
\tabnote{Sources and observation units vary by panel. A: AO Table C-13 district--fiscal-year cells for 89 federal districts, FY2010--FY2025, including FY2016; FJC case-level observations exclude recorded multidistrict-litigation cases. B: ACS one-year PUMS, person-weighted private employee counts for lawyers (occupation 2100) in a balanced 42-state panel, 2010--2024 excluding 2020. C: BEA personal consumption of legal services, Census QSS customer-type receipts, \emph{American Lawyer} Am Law 100 gross revenue, and Thomson Reuters billable hours; nominal legal-services revenue is deflated to 2017 dollars using the BLS producer price index, whereas consumption uses the BEA legal-services consumption index. D: annual and quarterly Thomson Reuters practice-hour growth and published Legal Agent Benchmark practice scores. E: Vault associate-rated technology and Chambers AI-integration tiers, with FY2024 Am Law financials for matched firms. Summary statistics describe the available cells; no regression, controls, or fixed effects are involved.}
\end{table}

\section{Empirical specifications}\label{sec:spec}

\paragraph{Self-representation.} We estimate changes in self-representation using a difference-in-differences design that compares non-prisoner cases with prisoner petitions across 89 federal judicial districts. For district $d$, group $g \in \{\text{non-prisoner}, \text{prisoner}\}$, and fiscal year $t$, with $y_{dgt}$ the share of cases with a self-represented party in points,
\begin{equation}
y_{dgt} = \alpha_{dg} + \lambda_{dt} + \sum_{\tau \ne 2022} \delta_\tau\, \mathbf 1[g = \text{non-prisoner}]\,\mathbf 1[t = \tau] + \varepsilon_{dgt}. \label{eq:es}
\end{equation}
The unit effects $\alpha_{dg}$ absorb each group's level in each district, and the district-by-year effects $\lambda_{dt}$ absorb every shock common to civilians and prisoners in a district---a clerk's-office recoding, a filing-fee change, a pandemic backlog, a judge's standing order on AI. The $\delta_\tau$ are the non-prisoner series' movement relative to the prisoner series, relative to FY2022, the last pre-release year. Cells are weighted by their FY2010--2017 mean size and standard errors are clustered by district (89 clusters); the joint test of the post-period coefficients uses a wild cluster bootstrap with the null imposed \citep{cameron2008}. Three restricted versions replace the year interactions with a single treated $\times$ post-FY2023 indicator, a linear ramp in years since FY2022, and an adoption-weighted dose equal to the fiscal-year mean of ChatGPT weekly active users normalized to FY2025 $=1$. A detrended version removes a non-prisoner-specific linear trend estimated on FY2010--2022 before computing the $\delta_\tau$.

\paragraph{Case types.} From the FJC records we compute the plaintiff pro se rate by nature-of-suit family and fiscal year and compare FY2024--2025 with the FY2010--2022 mean, grouping families by whether a litigant must draft a non-form complaint.

\paragraph{Employment, spending, and revenue.} For each national series we fit a log-linear trend on 2010--2019, report the percent deviation of each post-2020 year from the trend, and scale it by the standard deviation of the 2010--2019 residuals. For the in-house share we report two benchmarks, a trend on 2010--2019 and a trend on 2016--2022. The within-state comparison of employer groups estimates, for state $s$, group $g \in \{\text{in-house}, \text{law firm}\}$ and year $t$,
\begin{equation}
\ln E_{sgt} = \alpha_{sg} + \lambda_{st} + \sum_{\tau \ne 2019} \beta_\tau\, \mathbf 1[g = \inhouse]\,\mathbf 1[t=\tau] + u_{sgt}, \label{eq:acs}
\end{equation}
weighted by 2010--2017 cell size with errors clustered by state. The $\beta_\tau$ trace the in-house-to-law-firm employment ratio relative to 2019.

\paragraph{Sector adoption.} For client sector $k$ and year $t$, with $E_{kt}$ in-house lawyers and $A_{kt}$ the share of businesses in the sector using AI from the Census Business Trends and Outlook Survey (zero before the survey began in September 2023),
\begin{equation}
\ln E_{kt} = \alpha_k + \gamma_t + \beta A_{kt} + e_{kt}, \label{eq:sector}
\end{equation}
estimated on 19 sectors and 2010--2024, weighted by 2019 size, with sector-clustered errors and a wild bootstrap. A state-level version instruments the state adoption rate with a shift-share of 2019 industry mix and national sector adoption \citep{goldsmithpinkham2020}.

\section{Results}\label{sec:results}

\subsection{Self-representation increased most in drafting-intensive suits}\label{sec:prose}

\begin{figure}[tbp]\centering
\includegraphics[width=\textwidth]{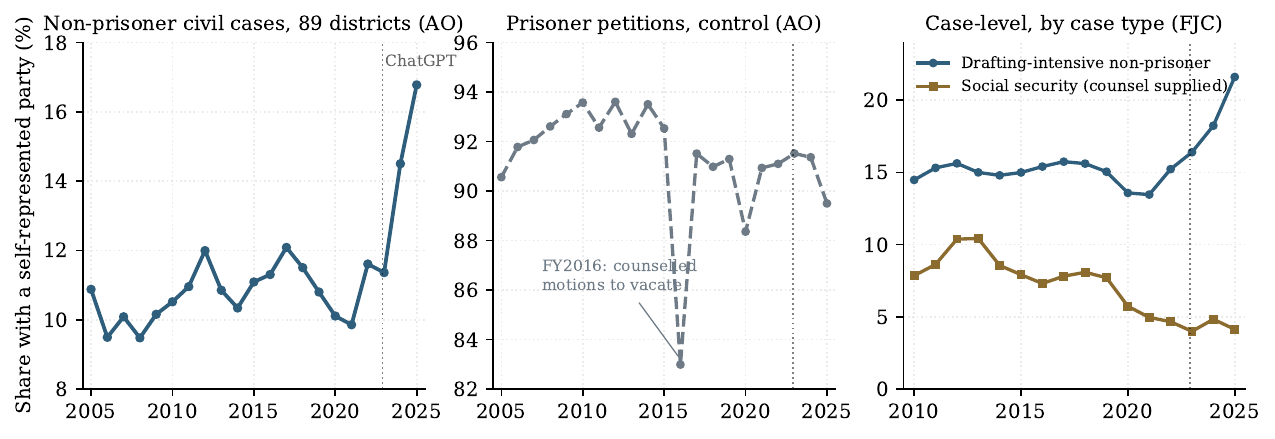}
\caption{Federal civil self-representation by litigant group and suit type, 2005--2025. Left and center: share of district-level civil filings with at least one self-represented party in non-prisoner cases and prisoner petitions, respectively, from Administrative Office (AO) Table C-13, FY2005--FY2025. District series exclude the Northern District of Florida and four territorial districts (89 districts). Right: share of Federal Judicial Center (FJC) Integrated Database civil records with a self-represented party, by selected suit family, FY2010--FY2025; records with recorded multidistrict-litigation status excluded. Fiscal years end September 30. The vertical marker denotes the November 2022 release of ChatGPT. Lines are descriptive, not fitted treatment effects.}\label{fig:prose}
\end{figure}

Figure~\ref{fig:prose} shows the two series. In the 89 main districts, the non-prisoner share of civil cases with a self-represented party averaged 10.7\% from 2005 to 2022 with a standard deviation of 0.8 points and never left the range 9.5--12.1\%. It was 11.6\% in FY2022, 11.4\% in FY2023, 14.5\% in FY2024, and 16.8\% in FY2025, 7.7 pre-period standard deviations above the mean. The prisoner series, 91--94\% throughout except for the documented 2016 wave, ended at 89.5\%. The rise is broad: ten of twelve circuits rose between FY2022 and FY2025, and in the 89 districts self-represented non-prisoner cases grew 86\% (22,234 to 41,377) while represented cases grew 21\% (169,308 to 205,232), so self-filed cases rose alongside a growing represented base; measured from FY2019, represented cases are 3\% lower and self-represented cases 62\% higher.

\begin{figure}[tbp]\centering
\includegraphics[width=0.85\textwidth]{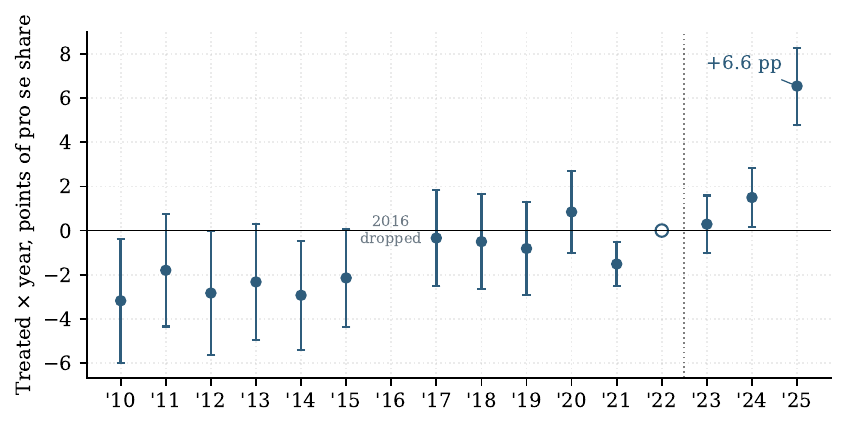}
\caption{Event study of the non-prisoner minus prisoner self-representation difference, FY2010--FY2025. Points are coefficients $\delta_t$ from equation~\eqref{eq:es}, relative to FY2022; whiskers are 95\% confidence intervals based on standard errors clustered by federal district (89 clusters). The outcome is the percentage of cases in an AO Table C-13 district--group--fiscal-year cell with at least one self-represented party. The regression includes district-by-group and district-by-year fixed effects, weights cells by their FY2010--FY2017 mean case count, and excludes FY2016. The sample includes 89 districts, non-prisoner cases and prisoner petitions.}\label{fig:es}
\end{figure}

Table~\ref{tab:courts} reports specification \eqref{eq:es}. The event-study coefficients for FY2023, FY2024, and FY2025 are $+0.3$, $+1.5$, and $+6.6$ points (SE 0.7, 0.7, 0.9), and the eleven pre-period coefficients range from $-3.2$ to $+0.8$ with a standard deviation of 1.3. The three post-period coefficients are jointly significant with a bootstrap $p$ below 0.001. The pre-period coefficients drift upward because the prisoner control falls about 0.2 points a year over the decade; removing a non-prisoner-specific trend fitted on FY2010--2022 leaves FY2025 at $+5.7$ points, 6.9 pre-period standard deviations, and $+0.41$ in log odds. Figure~\ref{fig:es} plots the path.

\begin{table}[tbp]\centering\caption{Civilian self-representation relative to prisoner petitions: event-study estimates on the district panel}\label{tab:courts}
\footnotesize
\resizebox{\textwidth}{!}{\begin{tabular}{lcccc}
\toprule
 & (1) & (2) & (3) & (4) \\
 & Share, weighted & Share, unweighted & Log odds, weighted & Share, FY2016 kept \\
\midrule
Treated $\times$ FY2019 & -0.81 & -1.77$^{*}$ & -0.03 & -0.81 \\
 & (1.07) & (0.92) & (0.09) & (1.07) \\
Treated $\times$ FY2020 & +0.84 & -0.31 & +0.13 & +0.84 \\
 & (0.94) & (0.92) & (0.08) & (0.94) \\
Treated $\times$ FY2021 & -1.52$^{***}$ & -2.21$^{***}$ & -0.12$^{**}$ & -1.52$^{***}$ \\
 & (0.51) & (0.77) & (0.06) & (0.51) \\
Treated $\times$ FY2023 & +0.29 & +0.41 & -0.05 & +0.29 \\
 & (0.66) & (0.72) & (0.07) & (0.66) \\
Treated $\times$ FY2024 & +1.50$^{**}$ & +1.51 & +0.04 & +1.50$^{**}$ \\
 & (0.67) & (0.98) & (0.06) & (0.67) \\
Treated $\times$ FY2025 & +6.55$^{***}$ & +8.63$^{***}$ & +0.52$^{***}$ & +6.55$^{***}$ \\
 & (0.89) & (1.43) & (0.08) & (0.89) \\
\midrule
Joint test, FY2023--25 $=0$, wild-bootstrap $p$ & $<0.001$ & $<0.001$ & $<0.001$ & $<0.001$ \\
Pre-period coefficient SD & 1.27 & 1.06 & 0.17 & 2.82 \\
FY2025 effect in pre-period SDs & 5.2 & 8.2 & 3.0 & 2.3 \\
Observations & 2,670 & 2,670 & 2,670 & 2,848 \\
District clusters & 89 & 89 & 89 & 89 \\
\bottomrule
\end{tabular}}
\tabnote{AO Table C-13 district--group--fiscal-year panel: 89 federal districts, non-prisoner civil cases (treated) and prisoner petitions (comparison), FY2010--FY2025. Entries are treated-by-year coefficients relative to FY2022 from equation~\eqref{eq:es}; parentheses contain district-clustered standard errors (89 clusters). Columns (1), (2), and (4) use the percentage of cases with any self-represented party as outcome (coefficients in percentage points); column (3) uses log odds. Columns (1), (3), and (4) weight each district--group cell by its average FY2010--FY2017 case count; column (2) is unweighted. Columns (1)--(3) omit FY2016; column (4) retains it. All specifications include district-by-group and district-by-fiscal-year fixed effects; there are no additional covariates. Joint post-period $p$-values use a wild cluster bootstrap. Significance symbols: $^{*}p<0.10$, $^{**}p<0.05$, $^{***}p<0.01$.}
\end{table}

The relative increase is concentrated in the later adoption years. Table~\ref{tab:lag} fits the three restricted lag structures. A single post-FY2023 step gives $+4.2$ points, averaging a year with no effect into two years with a growing one. A linear ramp gives $+2.3$ points a year. The adoption-weighted dose gives $+8.1$ points at FY2025 usage. The sequence matches the diffusion of the consumer tool---weekly active users went from roughly 40 million in FY2023 to 150 million in FY2024 and 490 million in FY2025---although this timing by itself cannot distinguish AI adoption from other gradual changes.

\begin{table}[tbp]\centering\caption{The effect grows with consumer adoption: restricted lag structures on the district panel}\label{tab:lag}
\footnotesize
\begin{tabular}{lcc}
\toprule
Specification & Coefficient & SE \\
\midrule
Step: treated $\times$ post-FY2023 & +4.24 & (1.00) \\
Linear ramp: treated $\times$ years since FY2022 & +2.29 & (0.42) \\
Adoption-weighted: treated $\times$ ChatGPT usage index & +8.11 & (1.24) \\
Event study, detrended, FY2025 & +5.67 & (0.89) \\
Event study, detrended, FY2025, log odds & +0.41 & (0.08) \\
\bottomrule
\end{tabular}
\tabnote{AO Table C-13 district--group--year panel, 89 districts, FY2010--FY2025 excluding FY2016. Each row estimates equation~\eqref{eq:es} with the full treated-by-year interactions replaced by the indicated treatment profile; the final two rows report FY2025 event-study coefficients after removing a non-prisoner-specific linear trend fitted on FY2010--FY2022. The outcome is the self-represented case share in percentage points except the final log-odds row. The step row uses a treated-by-FY2023--FY2025 indicator; the ramp is years after FY2022; adoption is interpolated fiscal-year ChatGPT weekly active users, normalized to 1 in FY2025 (0.08 in FY2023; 0.31 in FY2024). The adoption series is an exposure proxy, not litigant-level usage. All regressions include district-by-group and district-by-year fixed effects, use FY2010--FY2017 mean cell-size weights, and cluster errors by district; no additional controls.}
\end{table}

\begin{figure}[tbp]\centering
\includegraphics[width=0.8\textwidth]{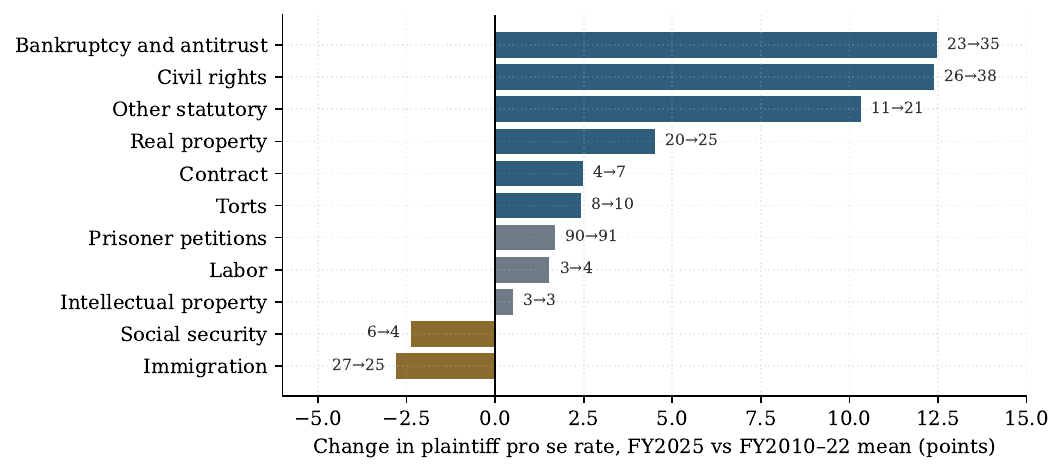}
\caption{Changes in plaintiff self-representation by nature of suit. Each bar is the FY2025 plaintiff pro se rate minus its FY2010--FY2022 mean, in percentage points, for civil cases in the FJC Integrated Database; numeric labels display the baseline and FY2025 rates. Suit families aggregate Administrative Office nature-of-suit codes as described in Appendix~\ref{app:data}. Cases with recorded multidistrict-litigation status are excluded. These are descriptive comparisons of case types, not regression estimates.}\label{fig:nos}
\end{figure}

Case-level heterogeneity is informative about the drafting-cost mechanism. Table~\ref{tab:nos} and Figure~\ref{fig:nos} compare the FY2025 plaintiff pro se rate with its FY2010--2022 mean by nature of suit. The rise is concentrated where a litigant has to write a complaint: civil rights from 26 to 38\%, other statutory actions from 11 to 21\%, real property from 20 to 25\%. It is small where representation is cheap or supplied by the market---intellectual property $+0.5$, labor $+1.5$, contract $+2.5$---and negative in social-security appeals, where fee-capped representatives handle most claims and the pro se rate fell from 6.4 to 4.0\%. Prisoner petitions moved 1.7 points on a 90\% base, and California small-claims filings, a forum in which lawyers may not appear and the pleading is a one-page form, show no rise at all (Appendix Figure~\ref{fig:ca}). Bankruptcy and antitrust actions, the family with the largest change at $+12.5$ points, rest on fewer than 3,000 filings a year. The one drafting-intensive family that did not rise relative to its mean, immigration, is dominated in FY2025 by alien-detainee habeas petitions filed during the enforcement surge, a policy shock rather than a technology one; its FY2022 to FY2025 change is $+17$ points.

\begin{table}[tbp]\centering\caption{Plaintiff pro se rate by nature of suit, FJC case-level records}\label{tab:nos}
\footnotesize
\resizebox{\textwidth}{!}{\begin{tabular}{lrrrrrr}
\toprule
Nature of suit & \multicolumn{3}{c}{Plaintiff pro se rate, \%} & Change & Records & Records \\
 & FY2010--22 & FY2024 & FY2025 & FY2025 $-$ pre & FY2025 & FY2010--22 avg. \\
\midrule
Bankruptcy and antitrust & 22.8 & 28.9 & 35.3 & +12.5 & 2,231 & 2,893 \\
Civil rights & 26.1 & 31.6 & 38.5 & +12.4 & 52,466 & 38,594 \\
Other statutory & 11.2 & 17.6 & 21.5 & +10.3 & 27,337 & 24,087 \\
Real property & 20.2 & 22.4 & 24.7 & +4.5 & 5,295 & 7,666 \\
Contract & 4.2 & 6.0 & 6.7 & +2.5 & 30,118 & 27,165 \\
Torts & 7.9 & 9.7 & 10.3 & +2.4 & 32,085 & 28,843 \\
Prisoner petitions & 89.8 & 90.5 & 91.5 & +1.7 & 51,579 & 54,566 \\
Labor & 2.9 & 3.6 & 4.4 & +1.5 & 12,754 & 17,261 \\
Forfeiture & 2.6 & 3.3 & 3.7 & +1.1 & 766 & 1,521 \\
Intellectual property & 2.9 & 3.0 & 3.4 & +0.5 & 18,909 & 14,146 \\
Social security & 6.4 & 4.6 & 4.0 & -2.4 & 15,637 & 17,954 \\
Immigration & 27.5 & 19.0 & 24.7 & -2.8 & 12,479 & 3,316 \\
\bottomrule
\end{tabular}}
\tabnote{Federal Judicial Center Integrated Database civil case-level records, FY2010--FY2025. The outcome is the percentage of records in a nature-of-suit family with a self-represented plaintiff, as recorded in FJC representation codes. Column 1 is the FY2010--FY2022 mean, columns 2--3 the FY2024 and FY2025 rates, and column 4 their FY2025-minus-pre-period difference in percentage points. Columns 5--6 report FY2025 record counts and annual average counts over FY2010--FY2022. Nature-of-suit families follow AO coding (Appendix~\ref{app:data}); cases with recorded multidistrict-litigation status are excluded. Estimates are descriptive unadjusted rates, with no controls or fixed effects.}
\end{table}

\subsection{Household spending on legal services did not fall}\label{sec:spend}

\begin{figure}[tbp]\centering
\includegraphics[width=\textwidth]{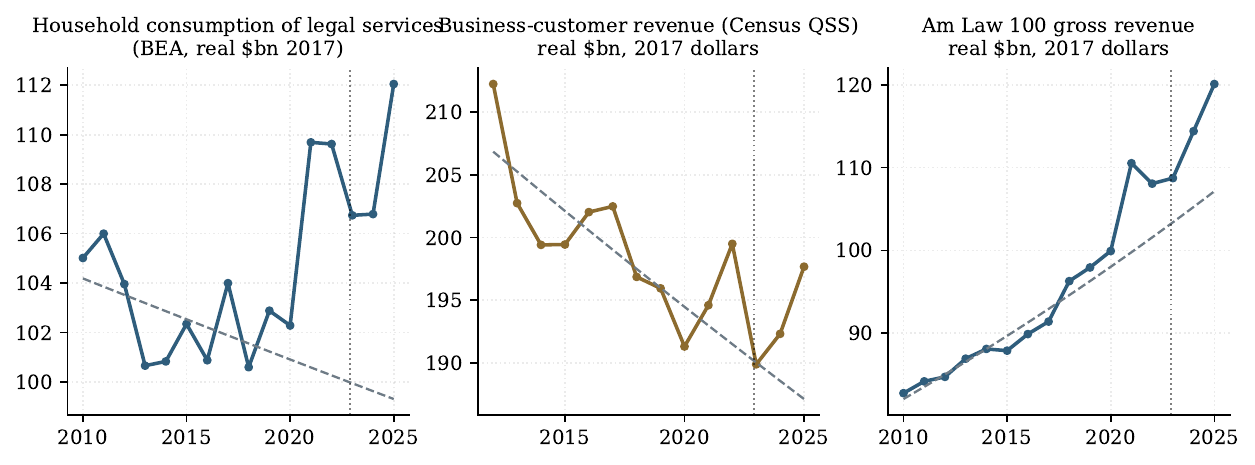}
\caption{Real demand for legal services relative to pre-2020 trends. Series show personal consumption of legal services (Bureau of Economic Analysis), law-firm revenue from business customers (Census Quarterly Services Survey), and Am Law 100 gross revenue (\emph{American Lawyer}). All amounts are expressed in 2017 dollars. Household consumption uses the BEA legal-services consumption price index; business-customer revenue and Am Law revenue use the Bureau of Labor Statistics legal-services producer price index. Dashed trends are log-linear fits to 2010--2019, except the business-customer series, fitted over 2012--2019. Plotted deviations are descriptive; no controls or fixed effects are estimated.}\label{fig:demand}
\end{figure}

Figure~\ref{fig:demand}, left panel, and Table~\ref{tab:demand} show real household consumption of legal services from the BEA. The series fell 0.3\% a year through 2010--2019 and moved above that trend in 2021, before generative models: 8.7\% above in 2021 and 8.9\% in 2022, then 6.6\% in 2023, 6.9\% in 2024, and 12.1\% in 2025, the largest deviation in the series at 6.9 pre-period standard deviations. The Census household-customer series, where published, agrees: 7.6\% above trend in 2022, its last year. The initial level shift precedes the spread of generative AI. During the adoption years, spending remained above its earlier trend while self-representation rose. That is the configuration the household block predicts---self-provision expanding activity rather than replacing paid activity---but these aggregate series cannot establish how much paid work individual self-represented litigants would otherwise have purchased. Am Law 100 real revenue follows the same shape, 10\% above trend in the 2021 deal boom and 11\% above in 2025 (Table~\ref{tab:demand}).

\begin{table}[tbp]\centering\caption{Deviations from 2010--19 trends: household spending, business revenue, and premium-firm revenue (percent)}\label{tab:demand}
\footnotesize
\begin{tabular}{lcccc}
\toprule
Series & 2010--19 trend (\%/yr) & 2023 & 2024 & 2025 \\
\midrule
Household consumption of legal services (BEA, real) & -0.32 & +6.6 & +6.9 & +12.1 \\
Law-firm revenue from business customers (QSS, real) & -0.77 & -0.1 & +2.0 & +5.5 \\
Am Law 100 gross revenue (real) & +1.78 & +5.0 & +8.3 & +11.4 \\
\bottomrule
\end{tabular}
\tabnote{Real annual series in 2017 dollars. Row 1: BEA personal consumption of legal services divided by its own legal-services price index; row 2: Census Quarterly Services Survey law-firm receipts from business customers deflated by the BLS legal-services PPI; row 3: \emph{American Lawyer} Am Law 100 gross revenue deflated by the same PPI. Column 1 is the annual slope of a log-linear trend fitted over 2010--2019 (2012--2019 for QSS); columns 2--4 are percentage deviations of observed real values in 2023, 2024, and 2025 from extrapolated trend levels. These are national descriptive time-series comparisons, with no covariates or fixed effects.}
\end{table}

\subsection{No clear acceleration in in-house lawyer employment}\label{sec:inhouse}

\begin{figure}[tbp]\centering
\includegraphics[width=\textwidth]{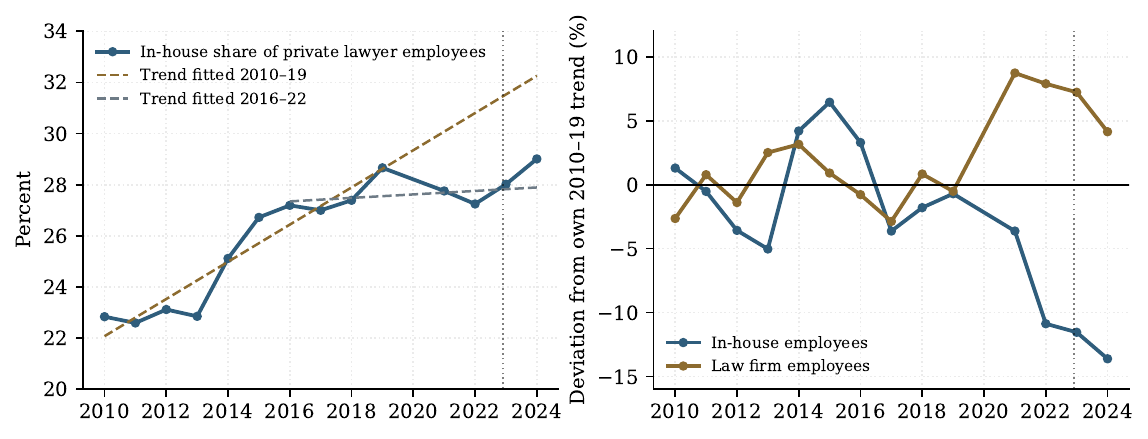}
\caption{In-house versus law-firm lawyer employment, 2010--2024. Left: the share of privately employed lawyers working outside the legal-services industry, with alternative linear trends fitted over 2010--2019 and 2016--2022. Right: percent deviations in each employer group's employment from its own 2010--2019 log-linear trend. Data are American Community Survey one-year public-use microdata, occupation 2100, private-sector employees in a balanced panel of 42 states; 2020 is omitted because no comparable ACS one-year PUMS was released. Person weights are used to estimate employment. These panels are descriptive and do not use regression fixed effects.}\label{fig:acs}
\end{figure}

Figure~\ref{fig:acs} and Table~\ref{tab:acs} show the location of lawyer employment. Among private-sector employee lawyers, the in-house share rose from 22.8\% in 2010 to 28.7\% in 2019 and reached 29.0\% in 2024. Most of this increase predates generative AI. The share rose sharply between 2013 and 2016, remained fairly flat through 2022, and then recovered from the temporary decline associated with the 2021--2022 law-firm hiring boom.

Trend comparisons reinforce this distinction between an old reallocation and a new break. The 2024 share lies about three percentage points below the extrapolation of the 2010--2019 trend, but about two points above the flatter 2016--2022 trend. In the within-state event study, the 2024 in-house-to-law-firm employment ratio is $0.04$ log points above 2019 (SE $0.05$); that difference is small relative to the cumulative increase during the 2010s. Separately, in-house lawyer employment was 13.6\% below its own 2010--2019 trend in 2024, while law-firm employment remained 4.1\% above its trend. Employment levels can lag task reallocation, but these estimates do not indicate an accelerated movement of lawyers to client firms through 2024.

\begin{table}[tbp]\centering\caption{Lawyer employment by employer: deviations from trend and the within-state comparison}\label{tab:acs}
\footnotesize
\begin{tabular}{lcccc}
\toprule
 & \multicolumn{2}{c}{Deviation from own 2010--19 trend (\%)} & \multicolumn{2}{c}{Within-state DiD, log employment} \\
\cmidrule(lr){2-3}\cmidrule(lr){4-5}
Year & In-house employees & Law-firm employees & Treated $\times$ year & (SE) \\
\midrule
2010 &  &  & -0.328 & (0.047) \\
2013 &  &  & -0.330 & (0.065) \\
2016 &  &  & -0.076 & (0.055) \\
2018 &  &  & -0.070 & (0.046) \\
2021 & -3.6 & +8.7 & -0.039 & (0.044) \\
2022 & -10.9 & +7.9 & -0.057 & (0.048) \\
2023 & -11.5 & +7.2 & -0.023 & (0.061) \\
2024 & -13.6 & +4.1 & +0.036 & (0.050) \\
\midrule
2010--19 trend (\% per year) & +5.50 & +1.64 & \multicolumn{2}{c}{base year 2019; 42 state clusters} \\
Pre-period residual SD (\%) & 3.98 & 2.13 & \multicolumn{2}{c}{unit and state $\times$ year FE} \\
\bottomrule
\end{tabular}
\tabnote{ACS one-year public-use microdata, 42 states with both lawyer employer groups observed in each included year, 2010--2024 excluding 2020; privately employed lawyers are occupation 2100, and in-house is defined as employment outside legal services (industry 7270). Columns (1)--(2) are national employment deviations (percent) from separate 2010--2019 log-linear trends for in-house and legal-services employees; these are descriptive. Column (3) reports treated (in-house) by year coefficients relative to 2019 from equation~\eqref{eq:acs}, with log person-weighted employment as the outcome; column (4) reports their standard errors, not a separate model. The regression includes state-by-employer-group and state-by-year fixed effects, weights state--group cells by their FY2010--FY2017 average size, and clusters standard errors by state (42 clusters); no other controls are included.}
\end{table}

Industry-level AI use offers another comparison. Using Census BTOS AI-use rates and ACS in-house lawyer employment, the sector panel estimates $\beta=-0.004$ per percentage point of adoption (SE $0.011$; wild-bootstrap $p=0.79$). Estimates range from $-0.004$ to $+0.003$ across the reported specifications (Appendix Table~\ref{tab:sector}). The 2024 deviations from each sector's earlier employment trend are also not significantly related to adoption (Spearman $0.31$, $p=0.21$). This is evidence about sector-wide AI use, not direct measurement of AI adoption by legal departments. It nevertheless suggests that the sectors adopting AI most rapidly have not yet experienced the largest relative increases in in-house lawyer employment.

\subsection{Premium prices rose while hours remained nearly flat}\label{sec:premium}

\begin{figure}[tbp]\centering
\includegraphics[width=\textwidth]{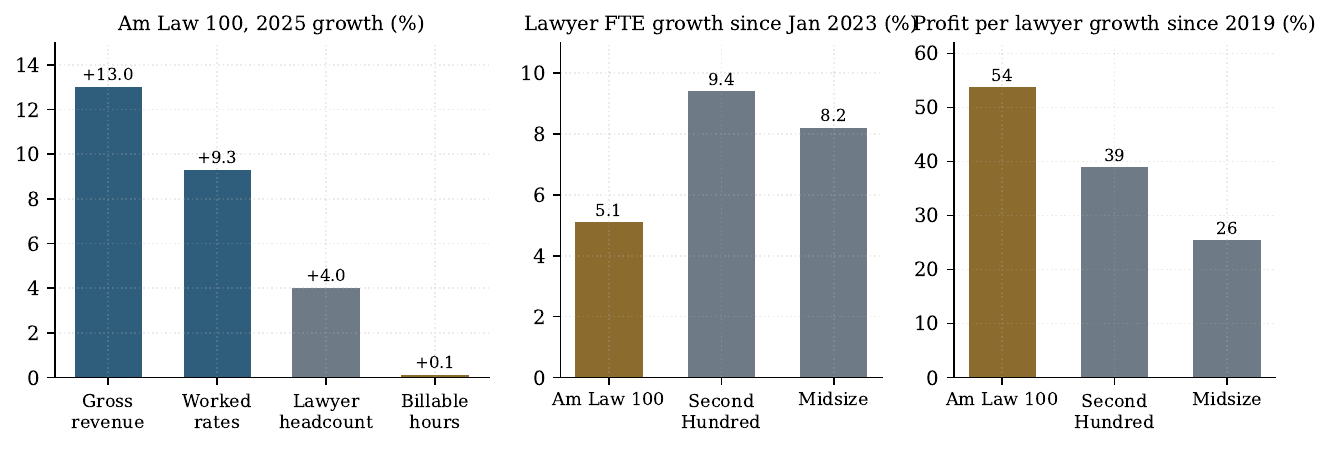}
\caption{Changes in price, volume, and staffing across US law-firm segments, 2025. Left: Am Law 100 gross-revenue growth from the \emph{American Lawyer} FY2025 survey versus billable-hour growth for the 50 Am Law 100 firms in Thomson Reuters' 2026 State of the US Legal Market sample, alongside Am Law 100 worked-rate growth from the same report (about 9.3\%; the all-segment average was 7.3\%). Center and right: Thomson Reuters comparisons of lawyer headcount, segment rates, and profitability for its surveyed firms. The Am Law revenue universe and Thomson Reuters hours sample differ; their percentages are descriptive comparisons and are not a firm-level accounting decomposition.}\label{fig:pv}
\end{figure}

The outside-provider market shows a different pattern. The Am Law 100 reported 13.0\% gross-revenue growth in 2025, while billable hours at the 50 Am Law 100 firms in Thomson Reuters' sample increased only 0.1\% (Figure~\ref{fig:pv}). Worked rates rose about 9\% at the Am Law 100, compared with 7.3\% across Thomson Reuters' broader sample. At the same time, large-firm hours overall grew 1.9\%, and midsize firms reported roughly 5\% hours growth in the second half of 2025. The largest firms thus saw particularly strong revenue growth without comparable growth in hours.

The relative prices make supplier substitution economically important. Standard hourly rates exceeded \$1,000 at the top segment, versus roughly \$600 at other firms. Thomson Reuters describes corporate clients moving more routine or moderately complex matters toward lower-priced firms \citep{tr2026,engelland2026}. The Am Law 100's real revenue was 11.4\% above its 2010--2019 trend in 2025 (Table~\ref{tab:demand}). These observations fit a composition channel in which premium providers retain a greater share of high-priced work, alongside a potential selection channel in which customers with weak outside options remain. A rise in average realized rates, however, cannot establish that the price of comparable work increased because of selection.

The samples measure different parts of the market. Am Law revenue includes non-hourly fees and offices outside the United States; Thomson Reuters' billable-hour series covers a subset of firms and matters. Equation~\eqref{eq:decomp} therefore provides an organizing identity rather than an exact numerical decomposition across the two sources. Corporate legal departments were already reallocating work across firms well before generative AI became widely available. The observed price-volume pattern is therefore evidence about a sourcing margin that AI could affect, not a dated causal estimate of AI's contribution to that reallocation.

\subsection{Practice differences largely predate generative AI}\label{sec:practices}

\begin{figure}[tbp]\centering
\includegraphics[width=\textwidth]{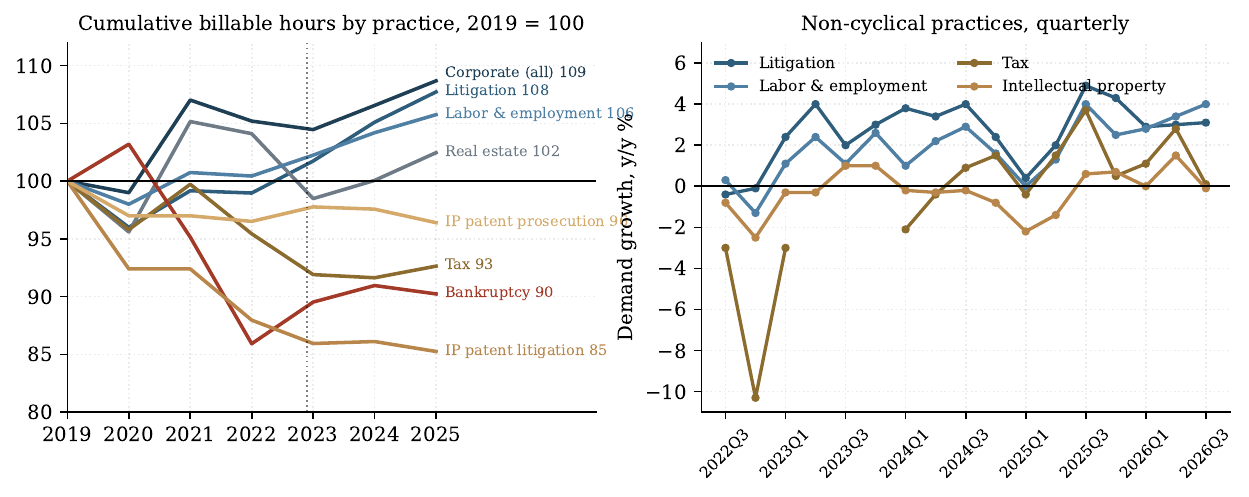}
\caption{Demand for outside legal work by practice. Left: index of billable hours at large US law firms (2019 $=100$), constructed by chaining annual practice-specific growth rates reported in Thomson Reuters' State of the US Legal Market, 2019--2025. Right: quarterly year-over-year billable-hours growth in litigation, labor and employment, tax, and intellectual property, from the Thomson Reuters Law Firm Financial Index, 2022Q3--2026Q3. Underlying samples vary between roughly 160 and 184 large US law firms by report; these are observed practice trends, not causal estimates of AI.}\label{fig:practices}
\end{figure}

Differences across legal practices help distinguish new technological change from longer-running demand shifts. Figure~\ref{fig:practices} chains Thomson Reuters' annual billable-hour growth by practice into indices with 2019 $=100$. By 2025, corporate, litigation, and labor-and-employment hours were 6--9\% above 2019, while tax, bankruptcy, and two patent categories were 4--15\% below. Much of this divergence was already present in 2022, before widely available generative AI.

Practice-level model scores do not establish that AI created the divergence. In one 2026 Legal Agent Benchmark evaluation, capability is relatively high in patent and tax work, but its Spearman correlation with the pre-AI 2022 hours index is $-0.65$ and its correlation with the 2022--2025 change is only $-0.36$ (Appendix Table~\ref{tab:practice}). A second benchmark, ContractBench, ranks commercial leases above IP licensing and commercial contracts on a different rubric (Appendix Table~\ref{tab:contractbench}). The reversal illustrates the problem with assigning an AI-exposure score to an entire practice: the relative difficulty depends on which tasks within the practice are evaluated.

Quarterly hours reinforce this reading. Litigation and labor-and-employment demand remained strong through 2026Q3; tax recovered after a period of weakness, while intellectual-property hours were broadly flat. Deal cycles, interest rates, client insourcing, and pricing changes can all affect these series. Their contribution to the paper is to show heterogeneity in the work supplied by law firms, including substantial differences that existed before consumer AI adoption.

\subsection{AI integration varies across firm market positions}\label{sec:firms}

The model also implies that AI incentives vary with a supplier's customer base. A firm competing for substitutable work has a strong reason to reduce the cost of delivery. A firm with relationship-specific or otherwise less substitutable demand may instead place more value on quality improvements. The firm-level evidence provides an initial comparison of these incentives.

\begin{figure}[tbp]\centering
\includegraphics[width=\textwidth]{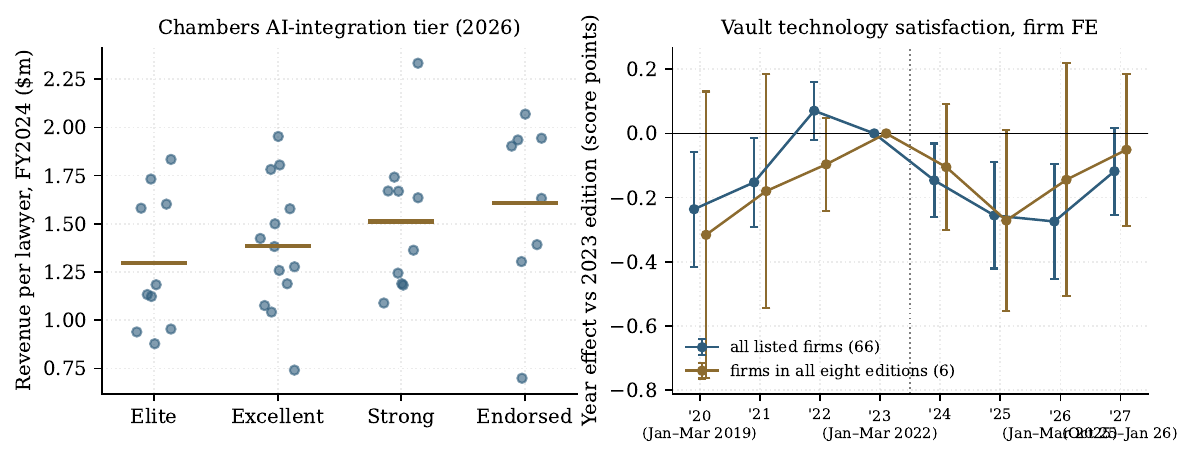}
\caption{AI integration, pricing position, and associate-rated technology. Left: mean FY2024 revenue per lawyer (\emph{American Lawyer}) by Chambers Associate 2026 AI-integration tier; 41 firms can be matched to financial data from the 64-firm Chambers survey. Right: estimated Vault Technology \& Innovation edition effects, relative to edition 2023, from a regression of the associate score on firm fixed effects and edition fixed effects; whiskers give 95\% intervals with standard errors clustered by firm. Vault covers its published top-30 firms per edition, editions 2020--2027; survey windows are marked on the axis. Neither measure directly estimates AI-related productivity.}\label{fig:firms}
\end{figure}

Among 41 firms with both Chambers Associate 2026 AI-integration tiers and Am Law financials, reported integration is stronger at firms with lower revenue per lawyer (Figure~\ref{fig:firms}). Elite-tier firms average \$1.30 million in revenue per lawyer versus \$1.61 million among Endorsed-tier firms; the Spearman correlation between a lower integration rank and higher revenue per lawyer is $0.32$ ($p=0.04$). Vault technology rankings are similarly concentrated in the stronger Chambers integration tiers: 60\% of Elite firms and 47\% of Excellent firms appear in Vault's technology top 30, compared with 16\% of Strong and none of the Endorsed firms.

These survey measures capture associates' assessments, not adoption intensity or cost savings. Nevertheless, the association is consistent with more urgent incentives to integrate AI among firms competing for contestable work. A separate Vault panel shows that associate-rated technology satisfaction recovered in late-2025 survey responses after declining in the preceding editions. Together, the findings place technology integration within the broader reallocation of legal demand rather than assuming adoption is uniform across professional-service suppliers.

\FloatBarrier
\section{Robustness}\label{sec:robust}

\paragraph{Alternative court samples and specifications.} The main court analysis excludes the Northern District of Florida because of an exceptional mass-tort filing episode. In the FJC case-level data, excluding cases with a recorded multidistrict-litigation status instead gives a non-prisoner pro se rate of 14.1\% in FY2022 and 20.6\% in FY2025, a 6.5-point increase. AO and FJC levels differ because the former counts filings and the latter includes records such as transfers and reopenings. Leaving out each of the 89 districts in turn puts the FY2025 event-study coefficient between 6.3 and 6.9 points. An unweighted specification yields 8.6 points, while the log-odds specification yields 0.52 (an odds ratio near 1.7). Keeping FY2016 largely affects pre-period dispersion because of its unusual prisoner-petition composition (Table~\ref{tab:courts}).

\paragraph{Pre-period trends and timing.} The prisoner and non-prisoner rates are not perfectly parallel before 2022. A non-prisoner-specific pre-period trend reduces the FY2025 coefficient from 6.6 to 5.7 points (Table~\ref{tab:lag}). Event-study coefficients in FY2020 and FY2021 are $+0.8$ and $-1.5$ points, within the pre-period range. Placebo three-year breaks dated 2008--2020 yield changes between $-1.7$ and $+2.4$ points, compared with $+4.6$ over FY2023--FY2025. These comparisons support the unusual size of the recent increase while leaving its precise causal attribution open.

\paragraph{Alternative employment definitions.} Adding self-employed lawyers and partners to both employment groups lowers the level of the in-house share to 19.8\% in 2019 and 21.2\% in 2024, but does not reveal a clear post-2022 acceleration. The series remains below its 2010--2019 projected trend and close to its 2016--2022 path. The lawyer occupation code is stable across ACS code vintages, although a 2013 industry recode coincides with an earlier level shift. Both trend benchmarks are therefore retained in Figure~\ref{fig:acs}.

\paragraph{Sector exposure and practice composition.} Census BTOS adoption measures AI use somewhere in the business, not necessarily in its legal department. The shift-share IV variant has an unweighted first-stage $F$ of 15.1, but state-year cells for in-house lawyers are small and the second-stage intervals are wide (Appendix Table~\ref{tab:sector}). The sector regression is best read as an association with a broad exposure proxy. Practice comparisons also reflect cyclical influences: M\&A, real estate, and bankruptcy respond strongly to credit and deal conditions, so the quarterly discussion focuses on litigation, labor and employment, tax, and intellectual property, while keeping cross-practice correlations descriptive.

\FloatBarrier
\section{Capability and timing}\label{sec:capability}

\begin{figure}[tbp]\centering
\includegraphics[width=0.9\textwidth]{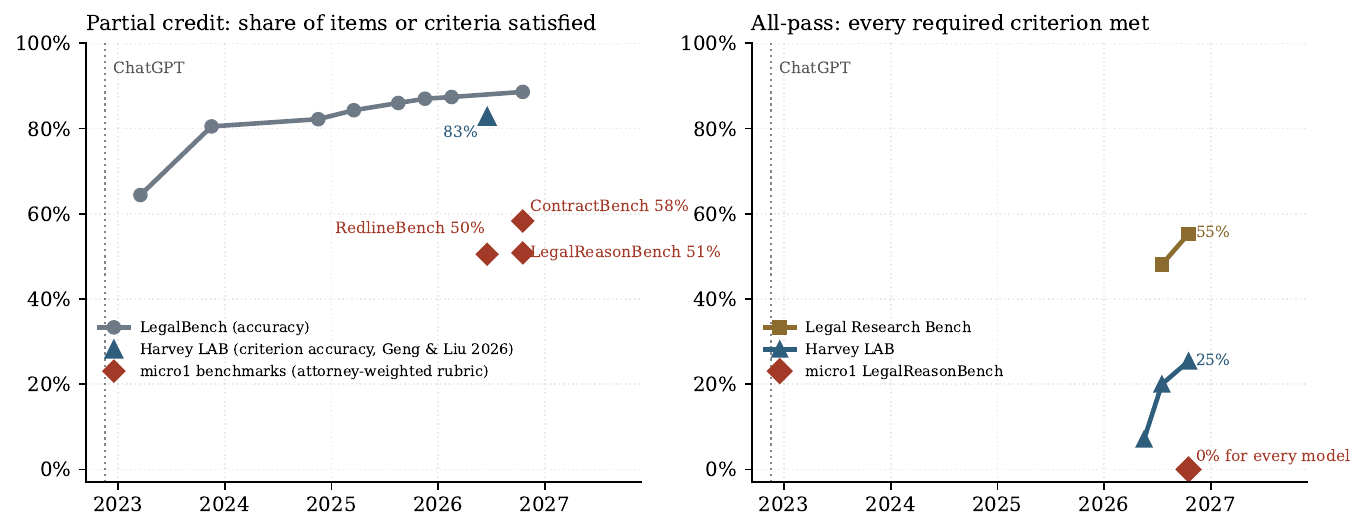}
\caption{Public measures of model capability on legal tasks, 2023--2026, by scoring rule. Left: partial-credit measures---LegalBench accuracy (Vals AI leaderboard, frozen snapshot accessed October 13, 2026, plotted at model release dates), Harvey Legal Agent Benchmark criterion accuracy for the best plain model in the independent full run of \citet{gengliu2026} (June 2026), and attorney-weighted rubric pass rates for ContractBench (October 20, 2026), LegalReasonBench (updated October 16, 2026), and turn-weighted scores for RedlineBench (June 17, 2026). Right: all-pass measures---Legal Research Bench and the Harvey Legal Agent Benchmark from Vals AI snapshots (July 2026 points as published in July; October points from the October 13 snapshot; the May 2026 Harvey point from the initial release), and LegalReasonBench, on which no model meets every criterion on any matter. Points use the best reported system at each date. Scores within a panel share a scoring rule but not a task set; scores across panels are not comparable.}
\label{fig:ladder}
\end{figure}

Consumer adoption and the reliability of professional AI tools are different objects. The household channel in Section~\ref{sec:model} depends on the accessibility of drafting assistance, proxied by the usage index in Table~\ref{tab:lag}. For enterprise clients, the economic effect of better tools depends on which categories of work they can perform or source more cheaply, as well as changes in outside-provider costs. Figure~\ref{fig:ladder} collects published benchmarks spanning legal reasoning, research, contract review, negotiation, and end-to-end matter completion, separated by scoring rule: the left panel shows partial-credit measures (share of items or criteria satisfied) and the right panel all-pass measures (share of tasks or matters on which every required criterion is met).

Legal-AI capabilities have improved substantially, but performance depends on task complexity and how success is measured. On LegalBench, the highest reported legal-reasoning accuracy rose from roughly 64\% in early 2023 to 89\% in October 2026. More demanding evaluations also show progress, although reliable completion remains limited. On Legal Research Bench, the all-pass rate increased from 48\% in July 2026 to 55\% in October. On Harvey's Legal Agent Benchmark, it rose from about 7\% in May to 25\% in October. Unlike accuracy or partial-credit measures, all-pass scores require every criterion for a task to be satisfied.

micro1's three expert-authored benchmarks further illustrate the challenges of completing complex legal work. The best models achieved 58\% on ContractBench and 51\% on LegalReasonBench under attorney-weighted rubrics, and 50\% on RedlineBench under a turn-weighted rubric. No evaluated model completed a LegalReasonBench matter in full. For comparison, an independent evaluation of Harvey's benchmark found that the best plain model satisfied about 83\% of individual criteria \citep{gengliu2026}. Because these benchmarks differ in task design, scoring methods, and evaluated models, their scores cannot be directly compared. Nevertheless, the results point to a consistent distinction: strong performance on individual legal tasks does not yet translate into reliable completion of demanding, multi-step legal work.

These benchmarks evaluate different tasks using different scoring rules and denominators. Their percentages are therefore not directly comparable, especially when some award partial credit while others require every criterion to pass. Nevertheless, the results reveal an important distinction between performing individual legal tasks and reliably completing complex legal work. This distinction helps explain the different patterns observed in Section \ref{sec:results}. 

For individuals, AI can lower the cost of drafting and filing a complaint without necessarily ensuring its quality. Self-representation has increased, but evidence that AI-assisted complaints face higher dismissal rates suggests that drafting assistance does not guarantee successful outcomes.

For enterprises, the make-or-buy decision in equation~\eqref{eq:threshold} depends on the relative costs of handling a recurring class of legal work.  AI can reduce internal costs by assisting lawyers with individual tasks, but replacing outside counsel also requires reliable execution, verification, and professional judgment. On Harvey's Legal Agent Benchmark, the leading system fully completes about one-quarter of matters; on micro1's LegalReasonBench, no evaluated model completes a matter in full. These limitations are consistent with the absence of a clear post-2022 shift toward in-house legal employment, although they do not establish its cause. The relevant measure of AI progress is therefore not a universal capability score, but whether AI can reliably support the completion of specific classes of recurring legal work at a lower cost. Improvements along this margin are most likely to affect firms' sourcing decisions.

\FloatBarrier
\section{Conclusion}\label{sec:conclusion}

Lowering the cost of specialized knowledge does not uniquely determine where knowledge work will be done. The outcome depends on which tasks clients can perform or source more cheaply, whether demand recurs often enough to sustain internal capacity, and how outside suppliers price against improving alternatives. A model that includes these features permits more self-provision, reallocation from premium to lower-priced suppliers, changes in the mix of work and clients remaining at premium firms, and little immediate movement in firm boundaries.

Legal services reveals why these distinctions matter. Federal civil self-representation rose sharply after a long stable period, particularly in several categories where drafting a pleading is important. Real household spending on legal services remained strong. By contrast, the substantial movement of lawyer employment away from law firms began before generative AI and shows no clear new acceleration through 2024. Among outside suppliers, premium-firm revenue and realized rates rose with nearly flat hours while lower-priced segments grew faster. These observations are compatible with changes in work composition and client selection, but do not identify the contribution of AI or distinguish a changing rate mix from within-service markup growth.

The broader implication extends beyond law. In consulting, accounting, engineering, and other knowledge-intensive services, lower task costs reshape organizations through clients' sourcing decisions and suppliers' competitive positions. Measuring AI's effect on knowledge work therefore requires tracking the composition of work, the selection of customers across suppliers, and changes in within-service prices, rather than treating specialist-industry employment or average billing rates as sufficient statistics.

\bibliographystyle{apalike}
\bibliography{refs}

\appendix
\section{Data construction}\label{app:data}
\paragraph{Courts.} AO Judicial Business Table C-13, fiscal years ending 30 September, transcribed from the published web tables (FY2005--2018, FY2020) and PDFs (FY2019, FY2021--2025). The treated series uses the non-prisoner columns and the control the prisoner columns. The 89-district main series drops N.D. Florida, Guam, the Northern Mariana Islands, Puerto Rico, and the Virgin Islands. FY2016 prisoner petitions: all-district totals 52,531 (FY2015), 73,725 (FY2016), 57,685 (FY2017), with non-pro-se prisoner petitions rising from 3,903 to 12,401 and back to 4,995; the Judicial Business report attributes a 350\% rise in motions to vacate to \emph{Welch}. FJC Integrated Database, civil, cumulative file snapshot of June 2026, filings with FILEDATE in FY2010--2025, 4,897,039 records; PROSE in $\{1,3\}$ defines plaintiff-side and $\{1,2,3\}$ any-party self-representation; recorded MDL status is a positive MDLDOCK or ORIGIN in $\{6,13\}$. Nature-of-suit families: civil rights 440--448; other statutory 375--376, 400, 430, 450, 470, 480--490, 890--899, 950; contract 110--196; real property 210--290; torts 310--385; labor 710--791; intellectual property 820--880; social security 861--865; immigration 460--465; prisoner 510--560; tax 870--871.
\paragraph{Employment.} ACS one-year PUMS 2010--2024. Occupation 2100 (lawyers, judges, magistrates, and other judicial workers) is identical in the 2010, 2014, and 2018 code lists; judges are removed by excluding public administration (industries 9370--9590) and, in the employee samples, by the class-of-worker restriction. Employees are class of worker 1--2; the robustness version adds 6--7. The balanced panel keeps the 42 states with a positive in-house cell in every year. Replicate weights give survey standard errors; the regressions use cluster-robust errors.
\paragraph{Spending and revenue.} BEA Table 2.4.5U line for legal services (nominal) and its price index (2017 $=100$). Census QSS, NAICS 5411 taxable revenue by class of customer, deflated by the BLS PPI for legal services (PCU54115411) rebased to the 2017 annual average. Am Law 100 gross revenue 2010--2025 as published by The American Lawyer, deflated by the same index.
\paragraph{Practices and capability.} Thomson Reuters practice growth from the State of the US Legal Market reports 2020--2026 (year to November); where two editions report the same year the later one is used. Quarterly growth from the Law Firm Financial Index executive reports, 2022Q3--2026Q3, with press-release figures for earlier quarters. Capability: \citet{gengliu2026}, Table 4, criterion accuracy for Claude Sonnet 4.6 in Claude Code; Thomson Reuters' corporate category is the mean of six LAB corporate areas and both patent practices map to LAB's intellectual-property area.
\paragraph{Firms.} Vault Technology \& Innovation, editions 2020--2027, public top-30 lists with survey windows as published; Chambers Associate, Top Law Firms for AI Integration 2026, fieldwork August 2025--January 2026; Am Law FY2024 financials matched by name (41 of 64 Chambers firms).

\section{Proof of Proposition~\ref{prop:price}}\label{app:proof}
Fix $x_H$ and write $u=p^*-m_H\geq0$, $h=s_Hx_H\geq0$, and
$\bar\lambda=(\lambda_BD_B+\lambda_FD_F)/(D_B+D_F)$.
The first-order condition $G=0$ gives
\[
 u=\frac{D_B+D_F-\lambda_FhD_F}
         {\lambda_BD_B+\lambda_FD_F}
 \leq\frac{1}{\bar\lambda}<\frac{1}{\lambda_B}.
\]
Thus $1-\lambda_Bu>0$. Since both demand groups have positive
mass, $G=0$ implies $1-\lambda_F(u+h)<0$. Therefore,
$G_{\xi}=-\lambda_FD_F[1-\lambda_F(u+h)]>0$.
At the stable local optimum $G_p<0$, so the implicit-function
theorem gives $\partial p^*/\partial\xi=-G_{\xi}/G_p>0$.
Finally, $D_{H,\xi}=-\lambda_FD_F<0$ and $D_{H,p}<0$;
therefore $dD_H/d\xi=D_{H,\xi}+D_{H,p}(\partial p^*/\partial\xi)<0$.
These statements hold locally along a stable interior pricing branch;
with discrete adoption or multiple pricing optima a global price jump
need not obey the local derivative.

\section{Additional tables}\label{app:tables}
\begin{table}[h]\centering\caption{Practice-area capability against billable hours at large firms}\label{tab:practice}
\footnotesize
\resizebox{\textwidth}{!}{\begin{tabular}{lrrrrrr}
\toprule
Practice & Capability (Sonnet 4.6) & Hours 2019 & Hours 2024--25 & Index 2022 & Index 2025 & $\Delta$ 2022--25 \\
\midrule
IP patent litigation & 85.7 & -3.0 & -0.4 & 88 & 85 & -2.7 \\
IP patent prosecution & 85.7 & -1.3 & -0.7 & 97 & 96 & -0.1 \\
Tax & 83.9 & -2.0 & +0.4 & 95 & 93 & -2.8 \\
M\&A & 83.8 &  & +0.9 &  &  &  \\
Litigation & 82.4 & +0.5 & +2.9 & 99 & 108 & +8.8 \\
Bankruptcy & 81.9 & +0.7 & +0.4 & 86 & 90 & +4.3 \\
Corporate (all) & 81.3 & +1.4 & +2.0 & 105 & 109 & +3.5 \\
Labor \& employment & 80.9 & +1.6 & +1.7 & 100 & 106 & +5.3 \\
Real estate & 80.2 & +1.3 & +2.0 & 104 & 102 & -1.6 \\
\midrule
\multicolumn{7}{l}{Spearman, capability vs index 2022: -0.65 ($p=0.08$); vs index 2025: -0.54; vs change 2022--25: -0.36; vs acceleration: +0.36 ($p=0.38$)} \\
\bottomrule
\end{tabular}}
\tabnote{Thomson Reuters annual practice-area billable-hours growth for large US law firms and published Harvey Legal Agent Benchmark practice-area criterion accuracy (Claude Sonnet 4.6 baseline, 2026). Capability is a within-benchmark percentage, not an estimated effect. The 2022 and 2025 hours indices cumulate annual reported growth from 2019 $=100$, and the change column is the percentage change in indexed hours between 2022 and 2025. Rank correlations exclude missing practice cells; associated $p$-values are descriptive. No controls or fixed effects. The different practice mixes and non-AI trends preclude a causal interpretation.}
\end{table}
\begin{table}[h]\centering\caption{Client-side AI adoption and in-house lawyer employment: sector panel and state shift-share}\label{tab:sector}
\footnotesize
\begin{tabular}{lcc}
\toprule
Specification & Coefficient & SE \\
\midrule
\multicolumn{3}{l}{\textit{Sector $\times$ year panel, log in-house lawyers on sector AI-use rate (pp)}} \\
Baseline, weighted, sector and year FE & -0.0037 & (0.0114) \\
Unweighted & +0.0011 & (0.0180) \\
Drop sector 54 & -0.0043 & (0.0155) \\
2023 exposure scaled to four months & +0.0020 & (0.0130) \\
Drop 2023 & +0.0030 & (0.0135) \\
All employees outside legal services & -0.0031 & (0.0105) \\
Falsification: in-house paralegals & -0.0004 & (0.0090) \\
\multicolumn{3}{l}{\textit{State $\times$ year shift-share, 2SLS, log in-house lawyers on state AI-use rate (pp)}} \\
Unweighted (first-stage $F$ = 15.1) & -0.0045 & (0.0391) \\
Weighted (first-stage $F$ = 3.0) & -0.0602 & (0.0607) \\
\bottomrule
\end{tabular}
\tabnote{Top panel: ACS sector-by-year employment, 19 client sectors, 2010--2024 excluding 2020; outcome log number of in-house lawyers. Explanatory variable is sector AI-use prevalence in percentage points from Census BTOS. Pre-survey observations are coded zero by construction, not measured non-adoption. Baseline uses sector and year fixed effects, 2019 sector employment weights, and standard errors clustered by sector; subsequent rows vary weighting, exclusions, exposure timing, the employee definition, or occupation (paralegal falsification). No other controls. Bottom panel: state-by-year IV estimates for 2019 and 2021--2024, instrumenting state AI use by 2019 industry employment shares interacted with national sector adoption; state and year fixed effects, standard errors clustered by state, unweighted or weighted as labeled. Reported first-stage $F$ values refer to the excluded instrument. The sector measure captures broad business AI use, not specifically AI deployed in legal departments.}
\end{table}
\begin{table}[h]\centering\caption{Ivo-micro1 ContractBench: weighted rubric pass rate by agreement type, percent, all turns}\label{tab:contractbench}
\footnotesize
\begin{tabular}{lrrrrr}
\toprule
Model & Commercial Lease & MSA & Marketing Agreement & IP Licensing & Research Partnership \\
\midrule
Opus 5.5 & 69.1 & 58.2 & 61.3 & 54.0 & 54.7 \\
Kimi K3 & 67.5 & 54.0 & 54.2 & 54.7 & 56.5 \\
Grok 4.7 & 67.0 & 58.2 & 56.6 & 56.8 & 57.6 \\
GLM-5.3 & 66.3 & 51.5 & 53.8 & 53.7 & 51.8 \\
Gemini 3.8 Flash & 64.5 & 51.7 & 57.1 & 52.9 & 50.5 \\
DeepSeek V4 Pro & 57.7 & 45.0 & 47.1 & 47.3 & 44.2 \\
GPT-6 Astra & 53.1 & 53.9 & 42.2 & 43.6 & 36.7 \\
\midrule
Seven-model mean & 63.6 & 53.2 & 53.2 & 51.9 & 50.3 \\
Best model & 69.1 & 58.2 & 61.3 & 56.8 & 57.6 \\
\bottomrule
\end{tabular}
\tabnote{Published Ivo--micro1 ContractBench task-level results (October 20, 2026), covering 350 expert-authored review tasks across five agreement types. Cells show weighted percentage of attorney rubric criteria passed for completed gradeable redlines, aggregating turns, with priority weights 3, 2, and 1. Rows are evaluated AI models; columns are agreement types, not regressions. Commercial leases are mapped to real estate, IP licensing to intellectual property, and MSA, marketing, and research-partnership agreements to corporate practice for the descriptive comparison in Section~\ref{sec:practices}. No regression controls or fixed effects.}
\end{table}
\begin{table}[h]\centering\caption{Chambers AI-integration tiers, Vault technology lists, and Am Law FY2024 financials}\label{tab:chambers}
\footnotesize
\resizebox{\textwidth}{!}{\begin{tabular}{lrrrrr}
\toprule
Tier & Firms & Share in Vault technology top 30 & Revenue per lawyer (\$m) & Profit per equity partner (\$m) & Lawyers \\
\midrule
Elite & 10 & 60\% & 1.30 & 3.40 & 2,030 \\
Excellent performer & 15 & 47\% & 1.39 & 3.78 & 1,652 \\
Strong performer & 25 & 16\% & 1.51 & 3.50 & 1,155 \\
Endorsed & 14 & 0\% & 1.61 & 4.82 & 1,410 \\
\midrule
\multicolumn{6}{l}{Spearman, tier (1 = Elite) vs revenue per lawyer: $+0.32$ ($p=0.04$, $n=41$); Fisher exact, Elite/Excellent vs rest $\times$ Vault top 30: $p=0.001$} \\
\bottomrule
\end{tabular}}
\tabnote{Chambers Associate's 2026 AI-integration tier classification includes 64 participating firms (first numeric column). The Vault column gives the share of firms in each tier also appearing in Vault's Technology \& Innovation top 30 in the relevant survey window. FY2024 revenue per lawyer and partner profit come from \emph{American Lawyer} and are available for 41 matched firms; other financial and headcount columns summarize reported available firms, not necessarily all 64. Tier ranking runs from Elite (1) to Endorsed (4). Rank correlation is calculated over matched firm observations; Fisher's test is a descriptive comparison of top-two tiers versus others. No fixed effects or regression controls are used.}
\end{table}
\begin{figure}[h]\centering
\includegraphics[width=0.6\textwidth]{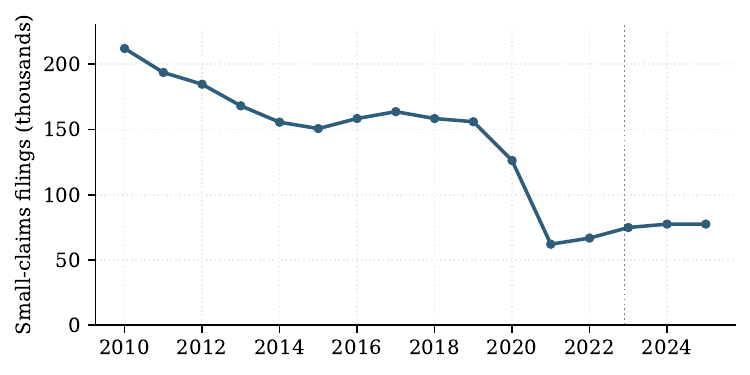}
\caption{California small-claims filings, fiscal years ending in June, 2010--2025. Data are published Judicial Council of California Court Statistics Reports. This venue uses a standardized short-form pleading and generally excludes attorney representation at the hearing. The plot is a descriptive comparison to the rise in federal self-representation, not an estimated control-group treatment effect.}\label{fig:ca}
\end{figure}

\end{document}